\documentclass[prc,showpacs,twocolumn,superscriptaddress]{revtex4}
\usepackage{graphicx}
\usepackage{dcolumn}
\usepackage{bm}

\begin{document}

\title{Microscopic description of fission in  odd-mass 
uranium and plutonium  nuclei  
with the Gogny 
energy density functional}

\author{R. Rodr\'{\i}guez-Guzm\'an}
\affiliation{ 
Physics Department, Kuwait University, Kuwait 13060, Kuwait.
}
\author{L. M. Robledo}
\affiliation{
Departamento  de F\'{\i}sica Te\'orica, 
Universidad Aut\'onoma de Madrid, 28049-Madrid, Spain}
\affiliation{Center for Computational Simulation,
Universidad Polit\'ecnica de Madrid,
Campus de Montegancedo, Boadilla del Monte, 28660-Madrid. Spain
}

\date{\today}

\begin{abstract}
The parametrization D1M of the Gogny energy density functional is used 
to study fission in the odd-mass Uranium and Plutonium isotopes with 
A=233,\ldots,249 within the framework of the Hartree-Fock-Bogoliubov 
(HFB) Equal Filling Approximation (EFA). Ground state quantum numbers 
and deformations, pairing energies, one-neutron separation energies, 
barrier heights and fission isomer excitation energies are given. 
Fission paths, collective masses and zero point rotational and 
vibrational quantum corrections are used to compute the systematic of 
the spontaneous fission half-lives t$_{SF}$, the masses and charges of 
the fission fragments as well as their intrinsic shapes. Although there 
exits a strong variance of the predicted fission rates with respect to 
the details involved in their computation, it is shown that both the 
specialization energy and the pairing quenching effects, taken into 
account fully variationally within the HFB-EFA blocking scheme, lead to 
larger spontaneous fission half-lives in odd-mass U and Pu nuclei as 
compared with the corresponding even-even neighbors. It is shown that 
modifications of a few percent in the strengths of the neutron and 
proton pairing fields can have a significant impact on the collective 
masses leading to uncertainties of several orders of magnitude in the 
predicted t$_{SF}$ values. Alpha-decay lifetimes have also been 
computed using a parametrization of the Viola-Seaborg formula. 
\end{abstract}


\maketitle

%
%
%

\section{Introduction.}

Fission is one of the many possible  decay modes of heavy   atomic  
nuclei \cite{Bjor,Specht} and, due to its characteristics, has 
attracted considerable attention since its discovery. 
It can be viewed \cite{Meitner} as the result of the competition between 
the nuclear surface energy, coming from 
the short range character of the strong nuclear interaction,
and the  Coulomb repulsion among protons. On the way to 
scission, atomic nuclei exhibit pronounced shape changes consequence of
the subtle balance between Coulomb, surface energy and quantum shell effects
associated to the underlying single particle structure of atomic nuclei.
Fission is then portrayed as a phenomenon where the shape of the nucleus,
described in terms of several deformation parameters,
evolves from the ground state to scission
\cite{Bjor,Specht,Wagemans,Baran-Kowal-others-review2015,SR-Review-2016}. 
How to account for those shape changes, and the associated quantum shell 
effects, still remains a major challenge in modern nuclear structure 
physics \cite{Baran-Kowal-others-review2015,SR-Review-2016} with an 
impact on both basic research and technology.

A better knowledge  of the fission process is required, for example, to 
deepen our understanding of the survival chances of a given element as 
one goes up in atomic number Z 
\cite{Baran-Kowal-others-review2015,Bender-1998,Warda-Egido-2012,%
Warda-Egido-Robledo-Pomorski-2002,Nature-Naza}, 
to account for the
competition between different decay modes 
(fission, $\alpha$-decay, cluster radioactivity, $\dots$) 
\cite{Viola-Seaborg,TDong2005,Rayner-UPRC-2014,%
Rayner-UEPJA-2014,Rayner-RaEPJA-2016,Rayner-No-odd,%
Robledo-Giulliani,Cluster-Warda} and to 
disentangle its role in  the r-process \cite{Panov,GMP_FR,syst-SH-SGR}.  
Furthermore,  fission is of high interest for the already existing and 
the new generation of nuclear reactors, the radioactive waste problem, 
weapon tests  and the production of super-heavy elements
(see, for example, \cite{Specht,Wagemans,Krappe,Sierk-PRC2015,JULIN-SHE}
and references therein).

Among the several theoretical frameworks used in   fission studies, the 
(constrained) mean-field  \cite{rs} approximation has emerged as a powerful 
tool. Here, the Hartree-Fock-Bogoliubov (HFB) method with constraints on
multipole moments, necking operators, etc is used to compute a multidimensional 
energy landscape defining the  potential energy to be used in fission dynamics
\cite{Baran-Kowal-others-review2015,SR-Review-2016}. 
Each configuration in the fission landscape
of a given even-even nucleus  is usually labeled by a set of shape deformation
parameters like the  quadrupole, octupole, $\dots${}
multipole moments, referred to collectively as {\bf{Q}}=(Q$_{20}$,Q$_{30}$, $\dots$). 
On the other hand, the mean-field framework also  
provides, in an unified manner, the collective inertias as well 
as  the zero-point quantum rotational and vibrational  energy
corrections \cite{Rayner-UPRC-2014} which are the required ingredients
to describe the quantum mechanical tunneling effect through the fission
barrier. With these  basic ingredients it is possible to make theoretical predictions 
about the spontaneous fission half-lives t$_{SF}$ and other 
relevant observable
\cite{SR-Review-2016,Rayner-UPRC-2014,Rayner-UEPJA-2014,%
Rayner-RaEPJA-2016,Robledo-Giulliani,Action-Rayner}. 
The mean field description of fission requires an effective 
energy density functional (EDF).
Popular choices are the non-relativistic Gogny
\cite{Rayner-UPRC-2014,Rayner-UEPJA-2014,Rayner-RaEPJA-2016,%
Action-Rayner,gogny-d1s,Delaroche-2006,Dubray,Younes-fission,%
Warda-Egido-Robledo-Pomorski-2002,Warda-Egido-2012}, 
Skyrme \cite{UNEDF1,Mcdonell-2,Erler2012,Baran-SF-2012,Baran-1981}, 
Barcelona-Catania-Paris-Madrid (BCPM) \cite{BCPM,Robledo-Giulliani} and 
relativistic \cite{Abusara-2010,Abu-2012-bheights,RMF-LU-2012,Kara-RMF}
EDFs.

In spite of the progress made in recent years in the study of the properties 
of odd-mass nuclei \cite{EFA-jsut,EFA-Bonneau,duguet-odd,EFA-Rayner-1,%
EFA-Rayner-2,EFA-Rayner-3,EFA-Rayner-4,Hamamoto-dd,reorient-1}, a task
greatly facilitated by developments in high-performance computing,  
microscopic fission studies in those nuclear systems are still rather scarce
within the EDF framework. This is mainly due to major technical difficulties 
\cite{Rayner-No-odd} that appear in the description of odd nuclei, as compared
to their even-even counterparts. 
First, to describe an odd-mass nucleus
time-reversal-breaking
one-quasiparticle "blocked" wave functions \cite{rs} should
be used. Therefore, time-odd
fields should be computed which increasing by a factor of two
the computing time required for the solution of the mean-field equations.
Second, several one-quasiparticle initial states have to be considered
for each configuration in the multidimensional energy landscape in order
to reach the lowest energy solution as the self-consistent character of 
the HFB equation does not guarantee to obtain the lowest-energy solution by blocking 
the lowest one-quasiparticle state.
Third, reorientation effects 
\cite{reorient-1,reorient-2} should also be taken into
account in the solution of the mean-field equations. Therefore, 
an approximation is required  to reduce the computational effort in
EDF  fission studies for heavy odd-mass systems. Within this context, 
the Equal Filling Approximation (EFA) 
\cite{EFA-jsut,EFA-Bonneau,duguet-odd,EFA-Rayner-1,EFA-Rayner-2,%
EFA-Rayner-3,EFA-Rayner-4,EFA-Decharge}
represents a reasonable alternative to the full fledged HFB plus blocking
procedure \cite{Rayner-No-odd}. 
The Hartree-Fock-Bogoliubov EFA (HFB-EFA) has already been formulated
in a fully Ritz-variational fashion \cite{EFA-jsut} by introducing a quantum statistical 
ensemble where the one-quasiparticle configuration $\mu$ to be "blocked" and its Kramer's 
partner ${\overline{\mu}}$ have the same probability (1/2). 
Under this assumption, time-reversal invariance is preserved and only time-even fields
have to be considered in the solution of the HFB equation. Let us stress that 
one of the main advantages of the  variational 
formulation of the HFB-EFA is that it allows the use of the standard gradient method 
\cite{Robledo-Bertsch2OGM} to solve the 
system of mean-field equations, with the subsequent simplification in the 
treatment of many  constraints at the same time
\cite{EFA-Rayner-1,EFA-Rayner-2,EFA-Rayner-3,EFA-Rayner-4}.

One of the most prominent experimental features in heavy odd-mass 
nuclei is their larger spontaneous fission half-lives as compared with 
their even-even counterparts \cite{Bjor,Specht,Holden-paper}. In order 
to explain such a feature two mechanisms  have been invoked in the 
literature, namely, the so called specialization energy 
\cite{special-Fong} and the quenching of pairing correlations. The 
specialization energy modifies the collective potential felt by the 
odd-mass nucleus on its way to scission making the inner fission 
barrier height higher than in the corresponding even-even case. It 
essentially arises from the assumption that the K quantum number, i.e., 
the projection of the angular momentum along the intrinsic nuclear 
axial symmetry axis, should be conserved along the fission process 
\cite{special-Fong,Rayner-No-odd}. On the other hand, in an odd-mass 
system, pairing correlations are quenched by the unpaired nucleon and 
there is a weakening of  the strength of the pairing field \cite{rs}. 
As the collective inertias exhibit a strong dependence on the inverse 
of the  square of the pairing gap 
\cite{Rayner-UPRC-2014,Rayner-UEPJA-2014,Rayner-RaEPJA-2016,%
Robledo-Giulliani,proportional-1,proportional-2} the weakening of 
pairing correlations in an odd system leads to a bigger collective 
inertia and therefore to an enhancement of the collective action. As a 
consequence, the spontaneous fission life time  t$_{SF}$ values take 
bigger values in the odd-A system than in the neighboring even-even 
nuclei \cite{Rayner-No-odd}.

Recently,  the fission properties of $^{251,253,255,257,259}$No 
\cite{Rayner-No-odd} have been evaluated within the HFB-EFA scheme. Our 
calculations  provide a reasonable  account of the ground state quantum 
numbers and deformations, pairing energies, one-neutron separation 
energies, excitation energies of  fission isomers as well as the inner 
and outer barrier heights. For those nuclei, we have also studied the 
systematic of the spontaneous fission and $\alpha$-decay lifetimes. 
Though there exists a strong variance of the predicted fission rates 
with respect to the details involved in their computation, it has been 
shown that  both the   specialization energy and the quenching of 
pairing correlations, taken into account selfconsistently within the 
HFB-EFA blocking procedure, lead to larger t$_{SF}$ values in odd-mass 
No isotopes  as compared with their even-even neighbors 
\cite{Holden-paper}.

In this paper we consider the fission properties of odd-mass 
neutron-rich uranium and plutonium nuclei within the (constrained) 
HFB-EFA \cite{EFA-jsut,EFA-Rayner-1,EFA-Rayner-2,EFA-Rayner-3,%
EFA-Rayner-4,Rayner-No-odd} based on the parametrization D1M \cite{gogny-d1m}
of the non-relativistic Gogny-EDF \cite{gogny}. Previous studies for even-even  
\cite{PRCQ2Q3-2012,Robledo-Rayner-JPG-2012,PTpaper-Rayner,Rayner-Robledo-JPG-2009}  
and also for odd-mass
\cite{Rayner-No-odd,EFA-Rayner-1,EFA-Rayner-2,EFA-Rayner-3,EFA-Rayner-4}
nuclei have shown that the parametrization D1M of the Gogny force preserves 
the predictive power of the well tested and well-performing
Gogny-D1S \cite{gogny-d1s} EDF while improving the description of nuclear masses
\cite{gogny-d1m}. In particular, previous  calculations 
\cite{Rayner-UPRC-2014,Rayner-UEPJA-2014,Rayner-RaEPJA-2016,Rayner-No-odd}
reveal that the Gogny-D1M EDF represents a reasonable starting point 
to describe fission properties in heavy and super-heavy nuclear systems.
In this work we have employed  the Gogny-D1M HFB-EFA, for the first time, 
to study the fission properties of odd-A U and Pu isotopes with A=233,\ldots,249
taken as illustrative samples. One should keep in mind
that a better knowledge of the fission properties of 
neutron-rich nuclei is required as these are the territories where the fate 
of the nucleosynthesis of heavy elements is  determined
\cite{Panov,GMP_FR,syst-SH-SGR}. 

The paper is organized as follows: in Sec.~\ref{Theory}, we briefly 
outline the HFB-EFA \cite{EFA-jsut} method as well as the methodology 
employed to obtain the one (1F) and two-fragment (2F) fission paths. We 
also discuss the relevant details regarding the computation of the 
spontaneous fission t$_{SF}$ and $\alpha$-decay t$_{\alpha}$ lifetimes. 
The results of our calculations are discussed in Sec.~\ref{RESULTS}. 
First, in Sec.~\ref{example-methodology}, we illustrate our methodology 
in the case of  $^{243}$U. The same methodology has been employed for 
all the other odd-mass nuclei considered in this work. The systematic 
of the fission paths, spontaneous fission half-lives as well as 
fragment mass and charge is presented in Sec.~\ref{syst-odd-mass-U-Pu}. 
In Sec.~\ref{pairstrength-odd-mass-U-Pu}, we discuss the impact of 
pairing correlations on the predicted spontaneous fission half-lives. 
To this end, we have also considered the Gogny-D1M EDF though with the 
strengths of the proton and neutron pairing fields increased by 5 and 
10 $\%$ to simulate effects of dynamical pairing that could appear as a 
consequence of symmetry restoration and/or a dynamical description of 
fission. For the sake of completeness, we will also include in the 
corresponding figures results already obtained for even-even U and Pu 
nuclei \cite{Rayner-UPRC-2014,Rayner-UEPJA-2014}. Finally, 
Sec.~\ref{Coclusions} is devoted to the concluding remarks and work 
perspectives.

\section{Theoretical framework} 
\label{Theory} 

In this section, we briefly outline the theoretical framework used in 
this study, i.e., the (constrained) HFB-EFA method used with the 
parametrization D1M  of the Gogny-EDF. The reader is referred to 
Ref.~\cite{EFA-jsut} for a  theoretical 
justification of the HFB-EFA based on ideas of quantum statistical 
mechanics.  We also describe the  methodology employed to obtain the fission paths 
in the studied odd-mass nuclei. Finally, we present the relevant 
details regarding the computation of the spontaneous fission t$_{SF}$ 
and $\alpha$-decay t$_{\alpha}$ lifetimes.

In the  HFB-EFA formalism the density matrix ${\rho}_{ij}^{(\mu,EFA)}$ 
and pairing tensor ${\kappa}_{ij}^{(\mu,EFA)}$ take the form
\begin{eqnarray} \label{rho-EFA}
{\rho}_{ij}^{(\mu,EFA)} &=& 
\left(V^{*}V^{T} \right)_{ij}
+\frac{1}{2}
\left(U_{i \mu} U_{j \mu}^{*} - V_{i \mu}^{*} V_{j \mu} \right)
\nonumber\\
&+&
\frac{1}{2}
\left(U_{i \overline{\mu}} U_{j \overline{\mu}}^{*} - V_{i \overline{\mu}}^{*} V_{j \overline{\mu}} \right)
\end{eqnarray}
and 
\begin{eqnarray} \label{kappa-EFA}
{\kappa}_{ij}^{(\mu,EFA)} &=& 
\left(V^{*}U^{T} \right)_{ij}
+\frac{1}{2}
\left(U_{i \mu} V_{j \mu}^{*} - V_{i \mu}^{*} U_{j \mu} \right)
\nonumber\\
&+&
\frac{1}{2}
\left(U_{i \overline{\mu}} V_{j \overline{\mu}}^{*} - V_{i \overline{\mu}}^{*} U_{j \overline{\mu}} \right)
\end{eqnarray}
where, the one-quasiparticle configuration, labeled by the index $\mu$, and
its Kramers' partner $\overline{\mu}$ have the same occupancies
n$_{\mu}$ = n$_{\overline{\mu}}$ = 1/2. 
In the above expression the $U$ and $V$ matrices represent the amplitudes of the Bogoliubov 
transformation, according to the standard notation of Ref \cite{rs}.
The above expressions of the density matrix and
pairing tensor are the ones of an statistical ensemble with two species 
(the blocked configurations $\mu$ and $\overline{\mu}$) both with probability $1/2$. 
Thanks to Gaudin's theorem \cite{Gaudin-th,Gaudin-Sara} (the equivalent
of Wick's theorem but for statistical ensembles), the
mean value of any observable in such a statistical ensemble takes the
same form as in standard HFB but replacing the densities by the ones of
Eqs (\ref{rho-EFA}) and (\ref{kappa-EFA}). This is the basic assumption
of the EFA.
Moreover, the Ritz variational principle applied to the total energy, 
written as a functional of both 
${\rho}_{ij}^{(\mu,EFA)}$ and ${\kappa}_{ij}^{(\mu,EFA)}$ 
leads to the standard HFB-EFA equation. As a consequence of its variational
nature the HFB-EFA equation can be solved using the 
successful gradient method \cite{Robledo-Bertsch2OGM}. 

In our calculations we assume axial symmetry and therefore we
can use the K quantum number (projection along the symmetry
axis of the angular momentum) to label the one-quasiparticle
excitations. On the other hand, reflection symmetry is allowed
to break as required by the physics of mass asymmetric fission.
As a consequence parity might not a good quantum number along
the whole fission path. The Gogny D1M EDF has been used in 
the calculation with the standard assumptions: the two body
kinetic energy correction is fully taken into account, 
the Coulomb exchange contribution to the energy is treated
in the Slater approximation \cite{CoulombSlater} and Coulomb antipairing is 
fully neglected. As reflection symmetry is broken, an
additional constraint on the center of mass is used
to avoid spurious center of mass effects 
\cite{PRCQ2Q3-2012,Robledo-Rayner-JPG-2012}. Finally,
the traditional zero-point rotational energy correction
$\Delta E_{ROT}=\langle \Delta \vec{J}^2\rangle / \mathcal{J}_{Yocc}$ as
well as the 
vibrational one have been added {\it{a posteriori}} to the HFB 
energies \cite{Rayner-UPRC-2014,Rayner-UEPJA-2014,Rayner-RaEPJA-2016}.

As mentioned before, reaching the lowest energy solutions
for each K quantum number requires the use of several
initial configurations. In the following we 
describe the  methodology employed in this study
to deal with this and other peculiarities of odd mass nuclei.
We will use $^{243}$U 
(see, Sec.~\ref{example-methodology}) as an illustrative 
example. 

{\it{Step 1)~Determination of the 1F and 2F solutions for the even-even 
neighbor $^{242}$U  within the constrained HFB framework}}. Those wave functions have been 
computed using the same methodology and (optimized) axially symmetric 
harmonic oscillator basis as in Ref.~\cite{Rayner-UPRC-2014}. We have 
employed constraints on the axially symmetric quadrupole  $\hat{Q}_{20}$ 
and octupole $\hat{Q}_{30}$ operators 
\cite{PRCQ2Q3-2012,Robledo-Rayner-JPG-2012} to obtain the   1F 
solutions. For sufficiently large quadrupole moments, 2F solutions 
have been reached by constraining on the necking operator 
$\hat{Q}_{Neck}(z_{0},C_{0})=\exp (-(z-z_{0})^2/C_0^2)$ used to
fix the number of particles in a region around $z_{0}$ of width $C_{0}$
\cite{Rayner-UPRC-2014,Rayner-UEPJA-2014,Rayner-RaEPJA-2016}.

\begin{figure*}
\includegraphics[width=0.46\textwidth]{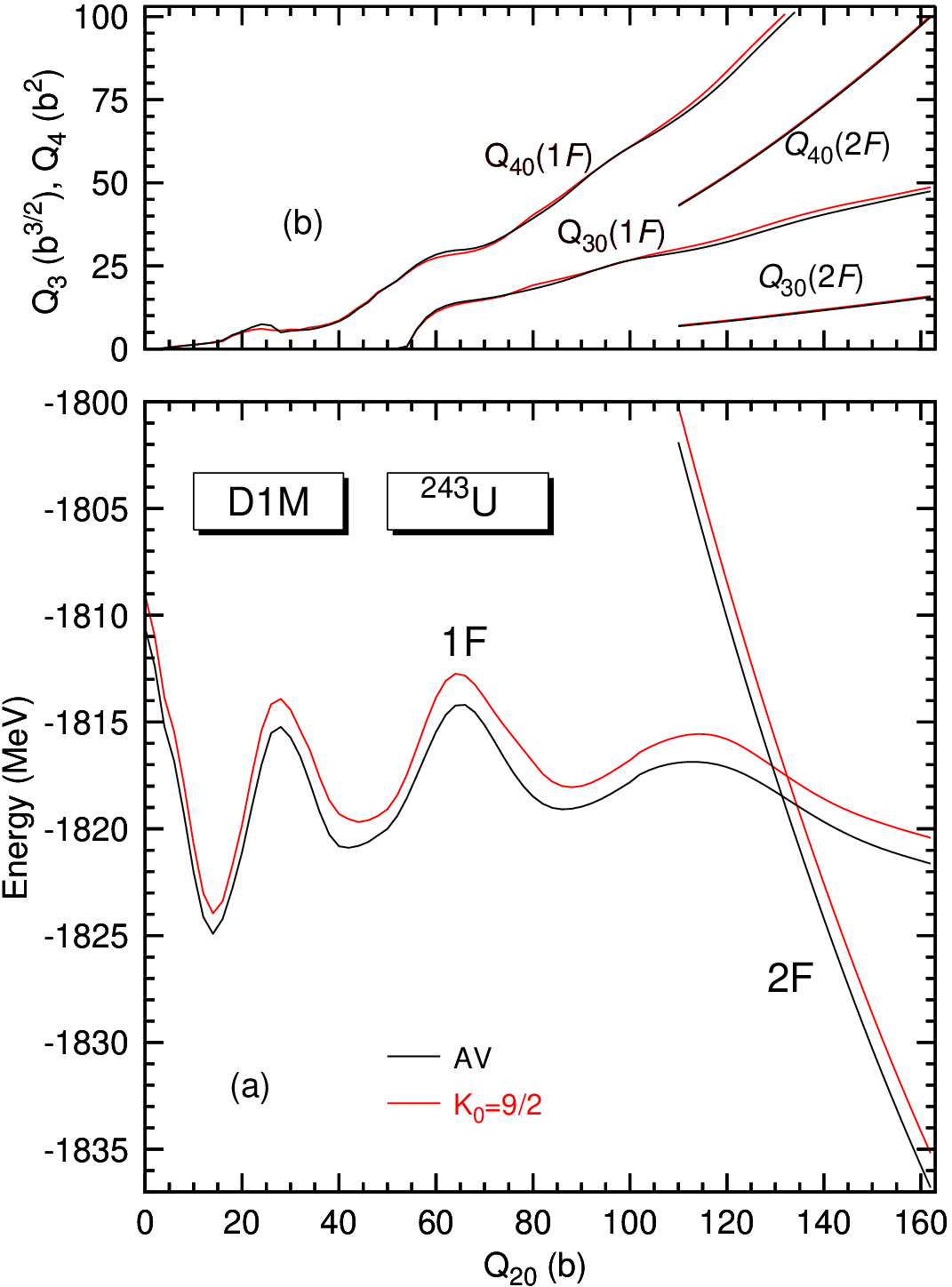}
\includegraphics[width=0.525\textwidth]{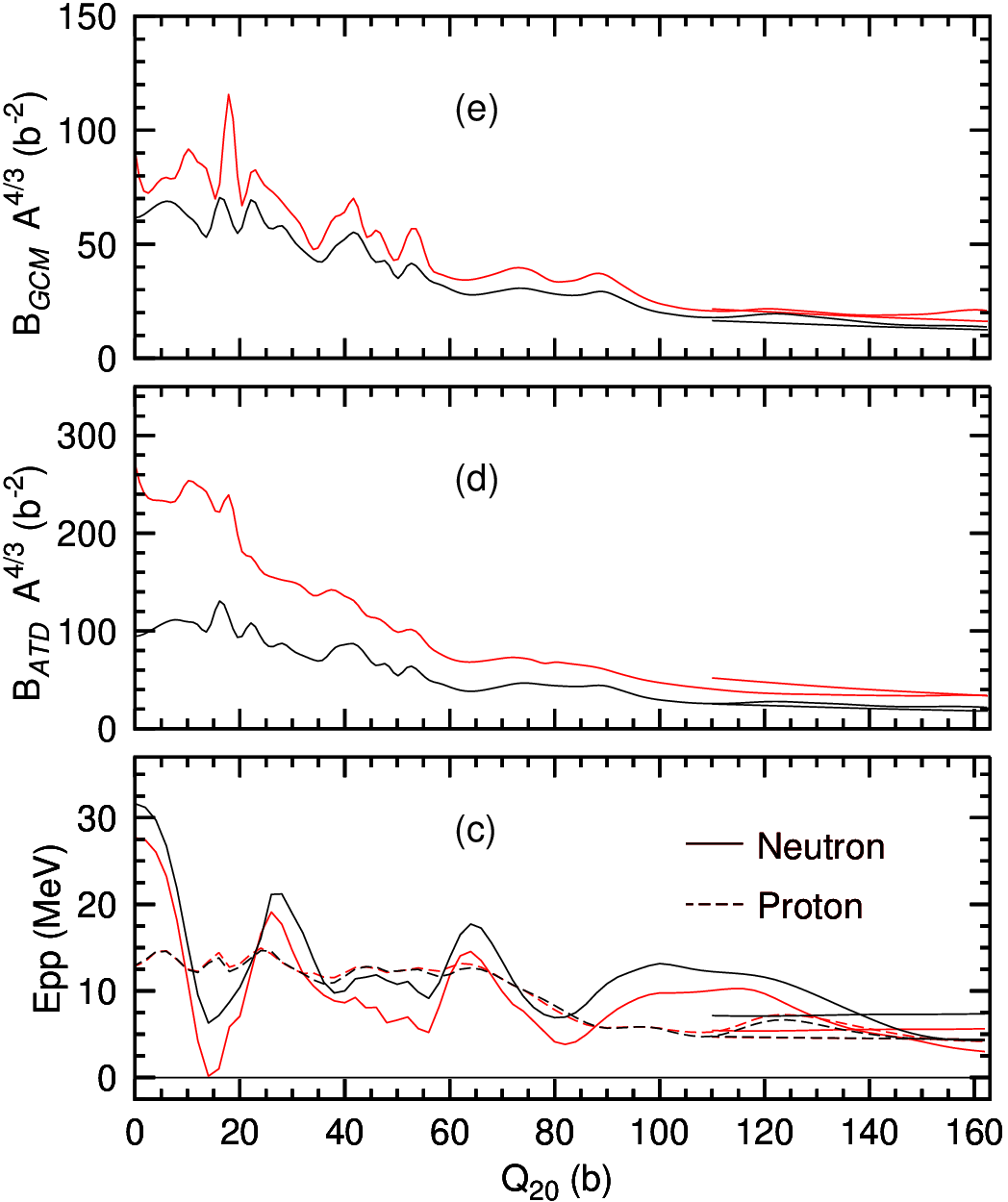}
\caption{(Color online) The K$_{0}$ = 9/2 HFB-EFA plus the zero point 
rotational energies, are plotted in panel (a) as functions of the quadrupole 
moment Q$_{20}$ for the nucleus $^{243}$U. Both the one (1F) and two-fragment 
(2F) solutions are included in the plot. The octupole Q$_{30}$ and hexadecapole 
Q$_{40}$ moments corresponding to the 1F and 2F solutions 
are shown in panel (b). The pairing interaction energies are depicted 
in panel (c) for protons (dashed lines) and neutrons (full lines). 
The collective ATD and GCM masses are plotted
in panels (d) and (e), respectively. Results corresponding to "average" 
(AV) HFB calculations for $^{243}$U have also been included in each panel. 
For more details, see the main text.  
}
\label{peda-1} 
\end{figure*}

{\it{Step 2)~Determination of the ``average"  1F and 2F solutions for 
$^{243}$U within the constrained HFB framework}}. We have used the 1F 
and 2F  solutions obtained for  $^{242}$U (Step 1) as initial wave 
functions to compute  "average" (AV) 1F and 2F solutions, respectively, 
for $^{243}$U (see, Fig.~\ref{peda-1}). Calculations have been carried 
out as for an even-even nucleus (i.e., no blocking is performed and all
wave functions with even number parity) but, 
with the mean value of the neutron number operator constrained to be 
$\langle \hat{N} \rangle$ = 151. We have employed the (optimized) HO 
basis resulting from Step 1. The zero-point quantum energy corrections have 
also been added {\it{a posteriori}} to the HFB energies.

{\it{Step 3)~Identification of the  (ground-state) K = K$_{0}$  quantum 
number for $^{243}$U within the  HFB-EFA}}. We have carried out HFB-EFA 
blocking calculations starting from the wave function corresponding to 
the absolute minimum of the AV 1F path (Step 2) in $^{243}$U. We have 
repeated the blocking procedure, using the same HO basis as in Step 2, 
several times so as to obtain five different solutions of the HFB-EFA 
equations for each of the values of K from 1/2 up to 11/2. Larger K 
values have not been  taken into account as the neutron single-particle 
levels corresponding to them are too far from the  Fermi surfaces. We 
have then identified, the K = K$_{0}$ quantum number corresponding to 
the lowest energy  (i.e., the ground state) among all the K-solutions 
obtained for $^{243}$U.

{\it{Step 4)~Determination of the 1F and 2F K$_{0}$-solutions for  
$^{243}$U within the constrained HFB-EFA}}. Having the corresponding  
AV 1F and 2F wave functions (Step 2) and the ground state quantum 
number K$_{0}$ for $^{243}$U, we have computed 1F and 2F 
K$_{0}$-solutions, respectively. Note, that we are assuming that the 
spontaneous fission of $^{243}$U will take place in a configuration 
with the same K = K$_{0}$ value as the one of the ground state  
\cite{Rayner-No-odd,special-Fong}. However, parity can be broken along 
the fission path. For each AV 1F and/or 2F state, we have repeated the 
blocking procedure, using the same HO basis as in Step 2, several times 
so as to obtain five different solutions of the HFB-EFA 
equations with the same K=K$_{0}$ value. This is the most time consuming
 step in our calculations as 
we move all over the AV 1F and 2F paths performing the required 
K$_{0}$-blocking for each {\bf{Q}}-configuration. This, already 
substantial, computational effort is greatly helped by the combined use 
of the HFB-EFA \cite{EFA-jsut} and the gradient method to solve its 
equations \cite{Robledo-Bertsch2OGM}. The 1F and 2F K$_{0}$-solutions 
with the lowest energy, for each ${\bf{Q}}$-configuration, are the ones  
used to build the ground state fission path for $^{243}$U (see, 
Fig.~\ref{peda-1}).

The rotational energy correction  $\Delta E_{ROT}$ to the HFB-EFA 
energies have been computed in terms of the Yoccoz moment of inertia 
using the formulas for even-even nuclei 
\cite{Rayner-No-odd,RCORR-1,RCORR-2,RCORR-3}. The reason for this choice is that an 
approximate angular momentum projection, like the one leading to the 
rotational energy correction, has not yet been carried out 
within the HFB-EFA. Work along these lines is in progress and will be 
reported elsewhere. On the other hand, previous finite temperature  
Adiabatic Time Dependent (ATD) results 
\cite{ATDHFB-T-1,ATDHFB-T-2,ATDHFB-T-3} can be extended to the HFB-EFA 
via its statistical density matrix operator \cite{EFA-jsut}. This, in 
turn, allows the computation of the ATD collective mass  and the 
zero-point vibrational energy correction $\Delta E_{vib}$ within the 
perturbative cranking approximation 
\cite{crankingAPPROX,Giannoni-1,Giannoni-2,Libert-1999,Rayner-No-odd}. 
Moreover, in this work we have also considered the alternative
GCM-like (perturbative)  collective masses and vibrational energy 
corrections \cite{Rayner-No-odd}. Though the expression for these quantities
lack a theoretical 
justification, as the one available in the ATD case, we have considered the 
GCM-like collective masses and vibrational energy corrections mostly to 
compare with the corresponding results already obtained for even-even U 
and Pu nuclei \cite{Rayner-UPRC-2014,Rayner-UEPJA-2014}.

\begin{figure*}
\includegraphics[width=0.3\textwidth]{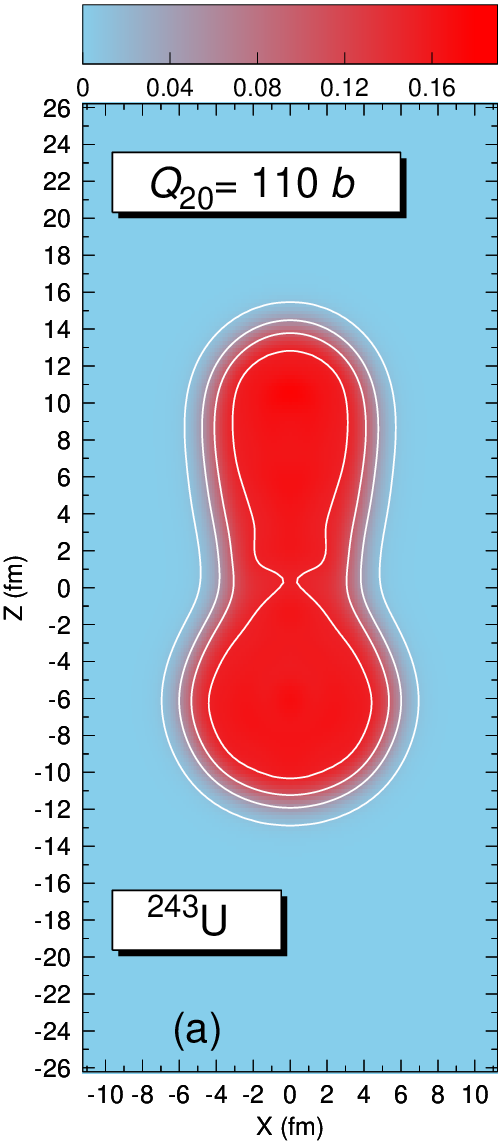}
\includegraphics[width=0.3\textwidth]{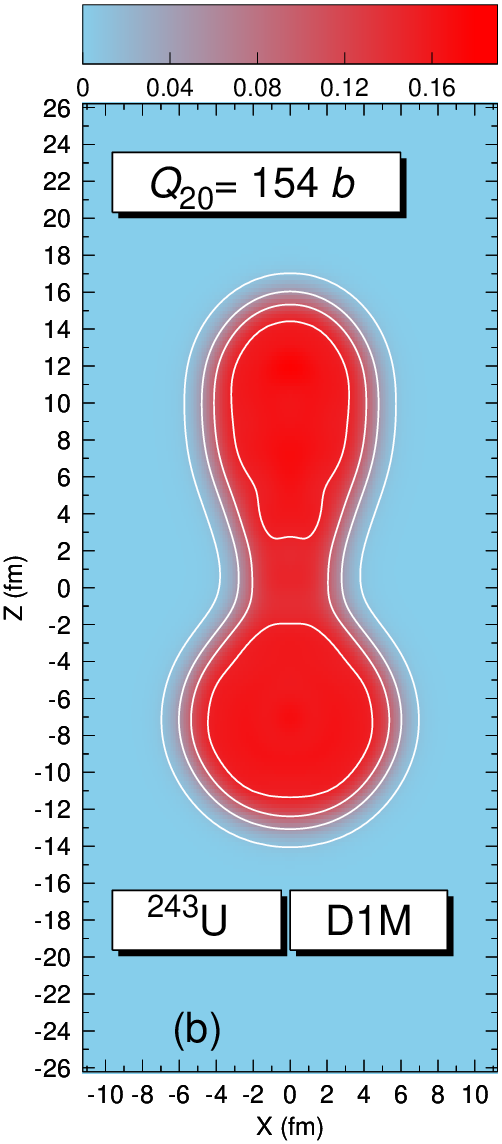}
\includegraphics[width=0.3\textwidth]{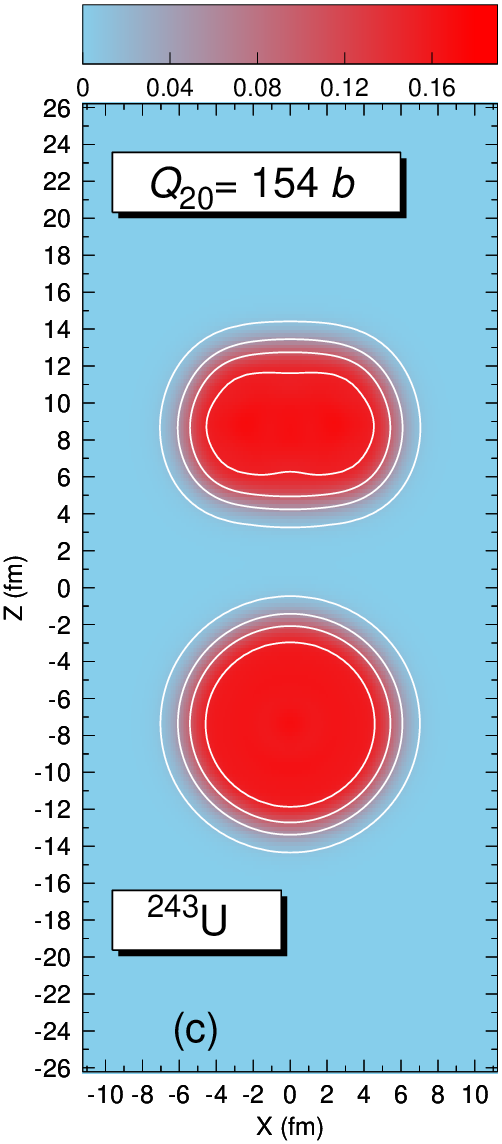}
\caption{(Color online) Density contour plots for the nucleus $^{243}$U at the 
quadrupole deformations $Q_{20}$=110 b [panel (a)] and  $Q_{20}$=154 b [panels (b) and (c)]. 
The
density profiles in panels (a) and (b) correspond to 1F configurations while the one 
in panel (c) represents a 2F solution. Densities are in units of fm$^{-3}$ and contour lines are drawn at 0.01, 0.05, 0.10 and 
0.15 fm$^{-3}$.
For more details, see the main text.  
}
\label{con-den-243U} 
\end{figure*}

We have obtained the spontaneous fission half-life t$_{SF}$ (in seconds)
within the Wentzel-Krammers-Brillouin (WKB) formalism  
\cite{Baran-TSF-1,Baran-TSF-2}

\begin{eqnarray} \label{TSF-WKB}
t_{SF} = 2.86 \times 10^{-21} \times \left(1 + e^{2S} \right)
\end{eqnarray}
where the action S along the fission K$_{0}$-path reads

\begin{small}
\begin{eqnarray} \label{action}
S = \int_{a}^{b} dQ_{20} \sqrt{2B(Q_{20})\Big[V(Q_{20}) - (E_{Min}+E_{0}) \Big] }
\end{eqnarray}
\end{small} 

In this work, the path to fission is determined by using the least energy
principle. This is a simplification over the alternative  approach of
considering the least action path.
In Eq. (\ref{action}), the integration limits a and b are the classical 
turning points for  $E_{Min}+E_{0}$. The energy $E_{Min}$ corresponds 
to the absolute minimum of the considered 1F K$_{0}$-path while $E_{0}$ 
accounts for the true ground state energy once quadrupole fluctuations 
are considered. The value of $E_{0}$ could be estimated using the 
curvature around the absolute minimum of the K$_{0}$-path and the 
values of the collective inertias 
\cite{Baran-SF-2012,Rayner-RaEPJA-2016}. However, we have followed the 
usual recipe \cite{Warda-Egido-Robledo-Pomorski-2002,Rayner-UPRC-2014} 
and considered $E_{0}$ as a free parameter that takes  four values 
(i.e., $E_{0}$ = 0.5, 1.0, 1.5 and 2.0 MeV). This allows us to estimate 
its impact on the predicted t$_{SF}$ values 
\cite{Robledo-Giulliani,Rayner-UPRC-2014,Rayner-UEPJA-2014,%
Rayner-RaEPJA-2016,Rayner-No-odd}. 
On the other hand, the collective potential $V(Q_{20})$ is given by the 
HFB-EFA energy corrected by the zero-point rotational $\Delta 
E_{ROT}$(Q$_{20}$) and vibrational $\Delta E_{vib}$(Q$_{20}$) energies. 
We have overlooked the  $E_{0}$-dependence of the prefactor in front of 
the exponential Eq. (\ref{TSF-WKB}) due to the large uncertainties in 
the estimation of the t$_{SF}$ values arising from other sources. 
Furthermore, in the computation of the t$_{SF}$ values 
Eq. (\ref{TSF-WKB}), the wiggles in the collective masses have been 
softened by means of a three point filter \cite{Rayner-UPRC-2014}.

Finally, in order to study the competition between the spontaneous 
fission and $\alpha$-decay modes, we have computed the t$_{\alpha}$ 
lifetimes using the Viola-Seaborg formula \cite{Viola-Seaborg} 
\begin{eqnarray} \label{VSeaborg-new}
\log_{10} t_{\alpha} =  \frac{AZ+B}{\sqrt{ {\cal{Q}}_{\alpha}}} + CZ+D + h_{log}
\end{eqnarray}
with parameters given in  \cite{TDong2005}. The ${\cal{Q}}_{\alpha}$ 
values (in MeV) are obtained from the calculated binding energies for 
U, Pu and Th nuclei. Within this context, the use of the Gogny-D1M EDF 
is particularly relevant as it provides a better description of the 
nuclear masses and it is expected to perform well in neutron-rich 
nuclei \cite{gogny-d1m}.

\section{Discussion of the results}
\label{RESULTS}

In this section, we discuss the results of our calculations for  
odd-mass U and Pu nuclei. First, in Sec.~\ref{example-methodology} we 
illustrate our methodology in the case of $^{243}$U. The systematic of 
the fission paths, spontaneous fission half-lives as well as fragment 
mass  and charge  is presented in Sec.~\ref{syst-odd-mass-U-Pu}. In 
Sec.~\ref{pairstrength-odd-mass-U-Pu}, we discuss the impact of pairing 
correlations on the predicted spontaneous fission half-lives. To this 
end, we have considered the Gogny-D1M EDF but with the strengths of the 
pairing fields increased by 5 and 10 $\%$, respectively.  For the sake 
of completeness, we also include in the corresponding figures results 
already obtained for even-even U and Pu nuclei 
\cite{Rayner-UPRC-2014,Rayner-UEPJA-2014}. 

\subsection{An illustrative example: The nucleus $^{243}$U}
\label{example-methodology}

\begin{figure*}
\includegraphics[width=1.00\textwidth]{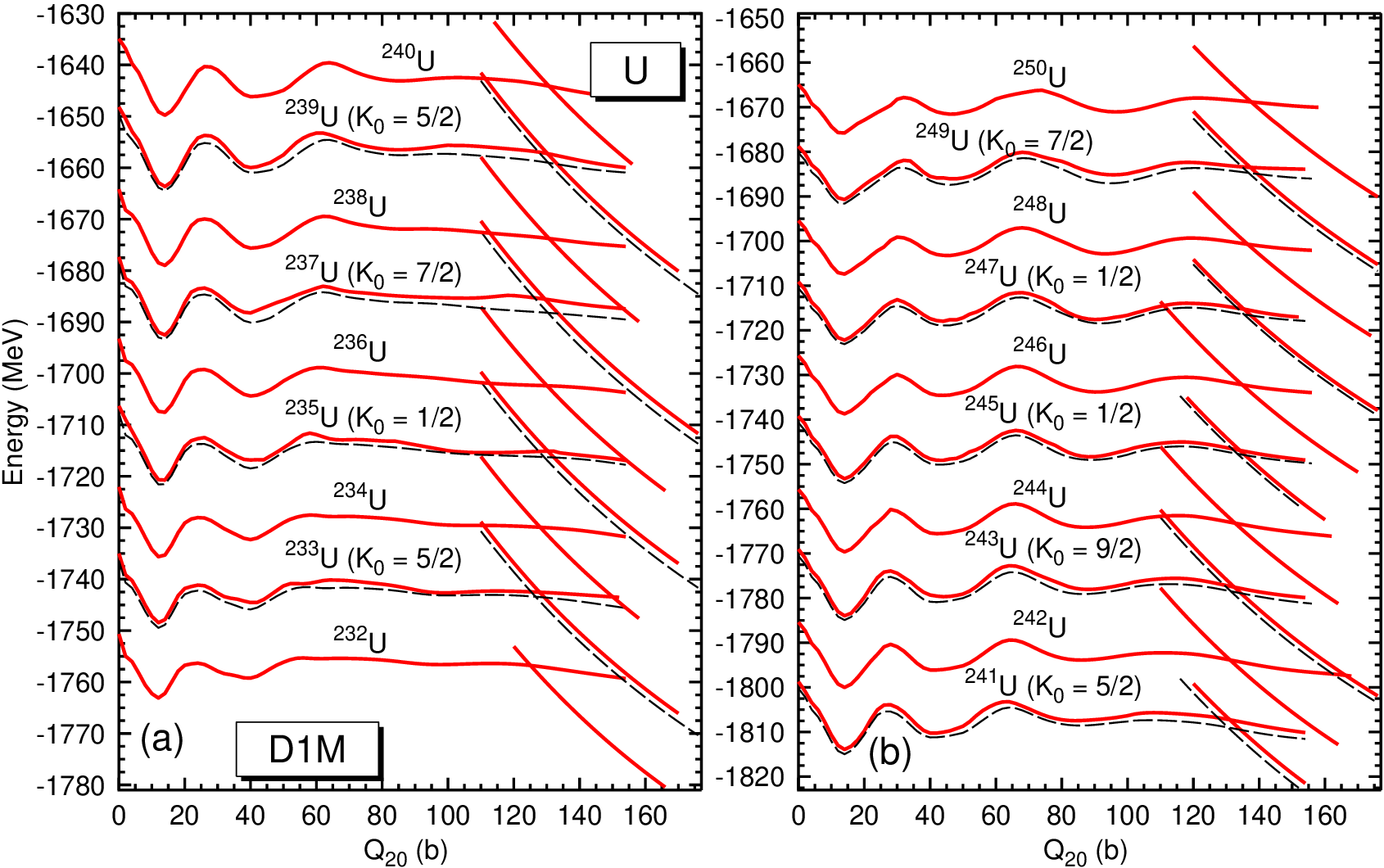}
\caption{(Color online) The ground state fission paths obtained for odd-mass U isotopes are plotted as 
functions of the quadrupole moment $Q_{20}$. The fission paths for  even-even isotopes 
are taken from Ref.~\cite{Rayner-UPRC-2014}. Starting from the nucleus $^{233}$U   ($^{242}$U)
in panel (a) [panel (b)], the curves have been shifted by 20 MeV in order to accommodate them in  a
single plot. For odd-mass nuclei, the corresponding K$_{0}$ values are included in the plot. The "average" (AV) fission
 paths for those odd-mass U nuclei are also depicted with dashed lines. For more details, see the main text. 
}
\label{bariers-U} 
\end{figure*}

In our calculations,  $^{243}$U is predicted to have a K$_{0}$ = 9/2
ground state with parity 
$\pi$ = -1 (i.e., a K$_{0}^{\pi}$ = 9/2$^{-}$ configuration). The 1F and 2F 
K$_{0}$ = 9/2 HFB-EFA 
plus the zero-point rotational 
energies, are plotted in panel (a) of 
Fig.~\ref{peda-1} as functions of the quadrupole moment Q$_{20}$. We are dropping
the parity of the ground state one-quasiparticle configuration to label the path
to fission because in this specific example, parity is broken at large quadrupole
deformations, beyond but near the fission isomer configuration. The zero-point 
vibrational energies $\Delta E_{vib}$ have not been included in the plot as
they are rather constant as functions of the quadrupole moment. However, we always consider 
such vibrational corrections in the computation of  the 
t$_{SF}$ and t$_{\alpha}$ 
lifetimes as well as other relevant quantities such as barrier heights, excitation energies 
of fission isomers, etc. The octupole Q$_{30}$ and hexadecapole Q$_{40}$ moments
are plotted in panel (b). In our calculations, an explicit constraint has not been included 
for  $\hat{Q}_{40}$ neither for other higher multipolarity operators. However, their 
average values are   automatically adjusted during the selfconsistent 
minimization of the 
HFB-EFA  energy.

As can be seen from the figure, the absolute minimum of the K$_{0}$ = 9/2 1F
path appears at $Q_{20}$ = 14b. The  first 9/2$^{-}$ fission isomer at 
$Q_{20}$ = 44b  lies 4.28 MeV above the ground state from which, it is 
separated by the inner barrier ($Q_{20}$ = 28b) with the height of 10.05 MeV.
Another noticeable feature from panel (a) is, the presence of a second octupole
deformed fission isomer 
at $Q_{20}$ = 88b that lies 5.91 MeV above the ground state. 
As we will see
later on (see, Sec.~\ref{syst-odd-mass-U-Pu}), second fission isomers are  
obtained for other odd-mass U and Pu nuclei. Those minima have already  
been found in previous Gogny-D1M calculations 
for even-even systems along both isotopic 
chains \cite{Rayner-UPRC-2014,Rayner-UEPJA-2014} as well as for other nuclei 
in this region of the nuclear 
chart \cite{Rayner-RaEPJA-2016}. This indicates that the shell effects 
\cite{Robledo-Giulliani,Kowal-3fi-1,Kowal-3fi-2,Delaroche-2006,Pas-3fi-1,Moeller-3fi-1,Mahrun-3fi-1,Berger-3fi-1,Zhao-3fi-1}
leading to those second fission isomers are systematically present in our 
mean-field  calculations. 
A second barrier 
with the height of 11.22 MeV is  found at $Q_{20}$ = 64b.
As can be seen from panel (b), both the second barrier
and the second fission isomer belong to the parity-breaking 
(i.e., $Q_{30}$ $\ne$ 0)
sector of the K$_{0}$ = 9/2 1F path in $^{243}$U. In fact, the left-right 
symmetry of the path is broken 
for $Q_{20}$ $\ge$ 54b. An outer barrier, with the height of 8.40 MeV, is also
visible at $Q_{20}$ = 112b. 

\begin{figure*}
\includegraphics[width=1.00\textwidth]{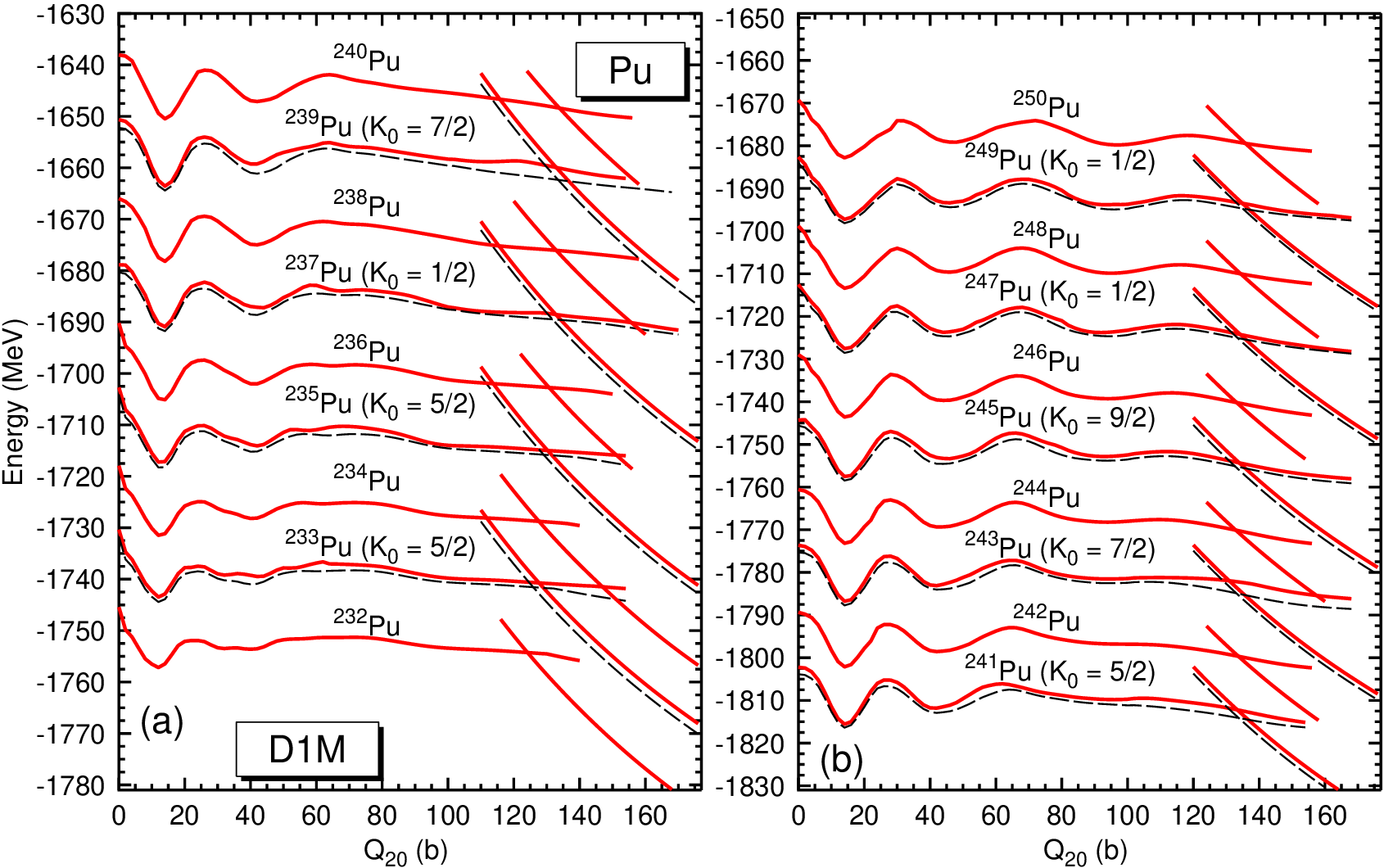}
\caption{(Color online) The ground state fission paths obtained for odd-mass Pu isotopes are plotted as 
functions of the quadrupole moment $Q_{20}$. The fission paths for  even-even isotopes 
are taken from Ref.~\cite{Rayner-UEPJA-2014}. Starting from the nucleus $^{233}$Pu   ($^{242}$Pu)
in panel (a) [panel (b)], the curves have been shifted by 20 MeV in order to accommodate them in  a
single plot. For odd-mass nuclei, the corresponding K$_{0}$ values are included in the plot. The "average" (AV) fission
 paths for those odd-mass Pu nuclei are also depicted with dashed lines. For more details, see the main text. 
}
\label{bariers-Pu} 
\end{figure*}

The previous values have been obtained with the K$_{0}$ = 9/2 configuration 
corresponding to the lowest energy for each quadrupole deformation (keep in mind 
that we assume the conservation of the K quantum number in the fission process).
Therefore, they
might or might not correspond to the lowest energy for a given $Q_{20}$. For
example, at the location of the first  fission isomer ($Q_{20}$ = 44b) 
the configuration with the lowest energy 
 corresponds to K = 5/2. On the other hand, at the location of
the second fission isomer ($Q_{20}$ = 88b) the configuration with the 
lowest energy corresponds
to K = 1/2.

In each panel of Fig.~\ref{peda-1}, we have also included results corresponding to AV HFB calculations 
for $^{243}$U. As can be seen from panel (a), the 1F and 2F HFB-EFA paths are always higher in 
energy than the AV ones. Moreover, the energy difference between the K$_{0}$ = 9/2 and AV paths is not
constant as a function of $Q_{20}$. This is a manifestation of the specialization energy effect partly 
due to the fact that we are following configurations with a fixed K$_{0}$ = 9/2 quantum
number. For example, the HFB-EFA inner, second and third barriers (10.05, 11.22 and 8.40 MeV)
are higher than the AV ones (9.68, 10.72 and 8.05 MeV). The same is also true for 
the excitation energies of the HFB-EFA first and second fission isomers (4.28 and 5.91 MeV)
when compared with the AV ones (4.02 and 5.82 MeV).

As can be seen from panels (a) and (b), not only 
the quadrupole but also the octupole and hexadecapole moments  
of the HFB-EFA and AV paths 
exhibit 
a rather similar behavior. This shows the very minute impact of blocking  
 in the mass moments characterizing the shape of $^{243}$U.
Panel (b) also reveals  that the $Q_{30}$ and $Q_{40}$ moments corresponding 
to the 1F [i.e., $Q_{30}(1F)$ and $Q_{40}(1F)$] and 2F [i.e., $Q_{30}(2F)$ and $Q_{40}(2F)$]
paths are rather different due to their separation  in the multidimensional
space of deformations.

\begin{figure*}
\includegraphics[width=0.49\textwidth]{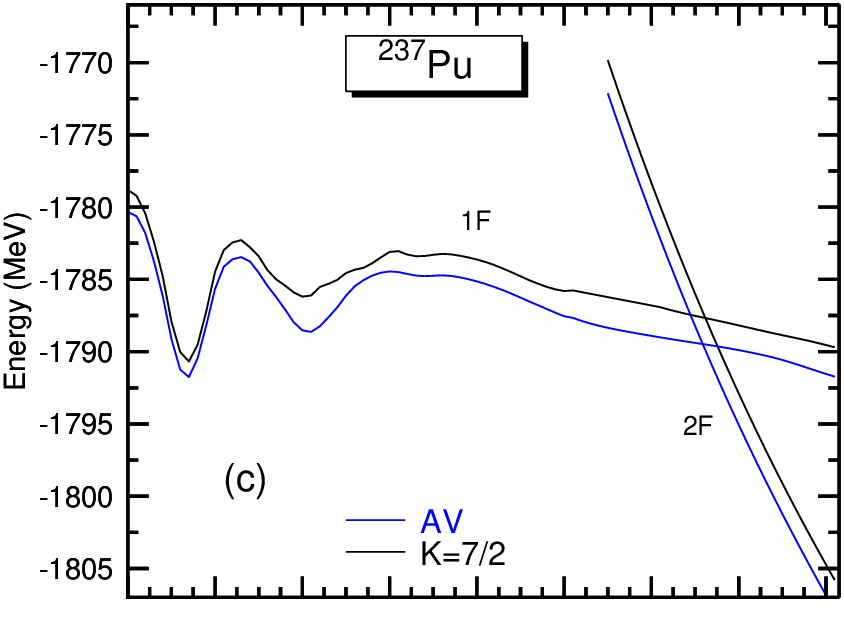} 
\includegraphics[width=0.49\textwidth]{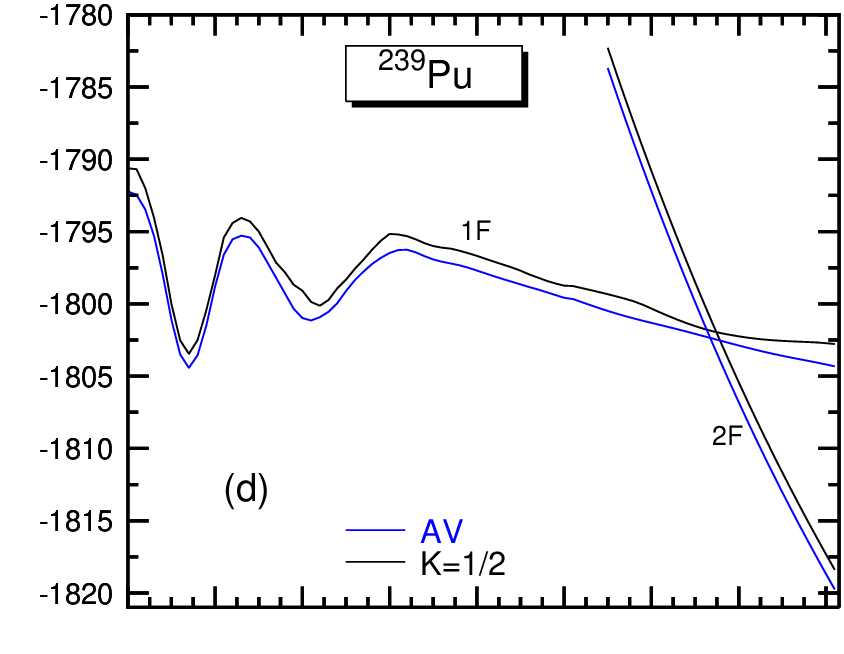} \\
\includegraphics[width=0.49\textwidth]{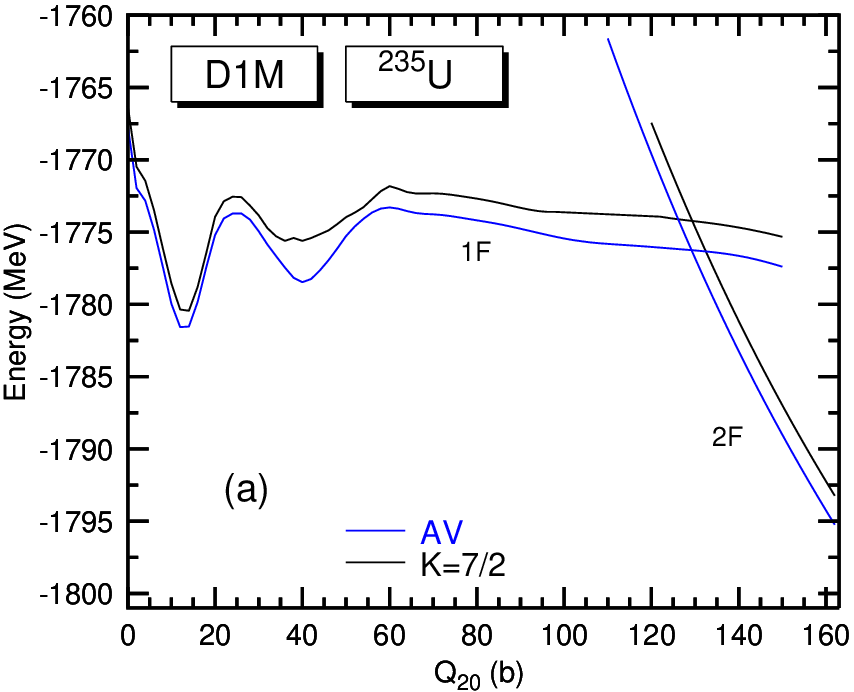} 
\includegraphics[width=0.49\textwidth]{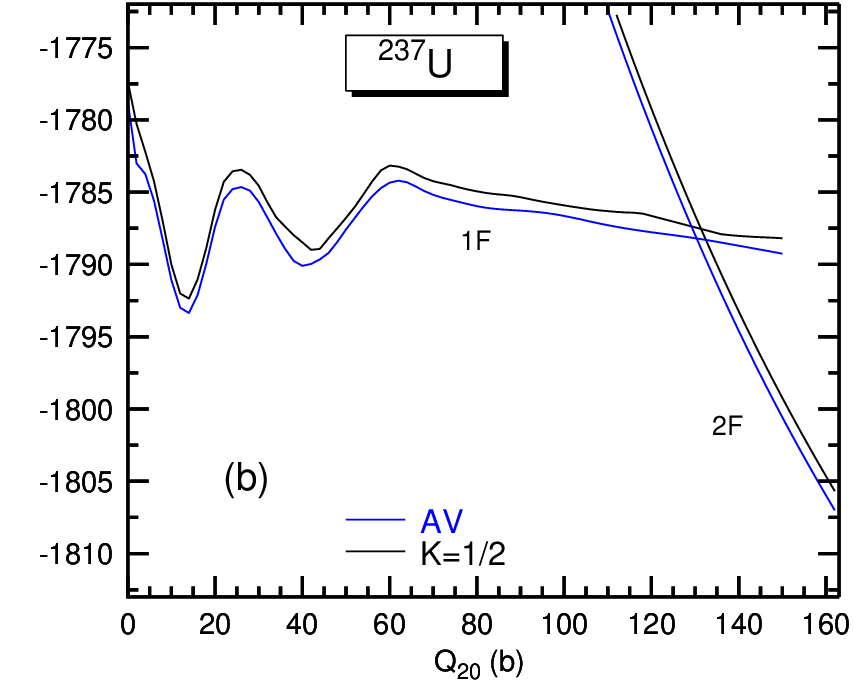}
\caption{(Color online) The K = 7/2 and 1/2 
HFB-EFA plus the zero point rotational energies, are plotted in panels
(a) [(c)] and (b) [(d)] for the nuclei  $^{235}$U and $^{237}$U
($^{237}$Pu and $^{239}$Pu)
as functions of the quadrupole moment Q$_{20}$. Both the 1F and 
2F solutions are included in the plots.
Results corresponding to "average" (AV) HFB calculations for 
$^{235}$U, $^{237}$U, $^{237}$Pu and $^{239}$Pu
have also been included in the panels. For more details, see the main text. 
}
\label{bariers-235U237U237Pu239Pu} 
\end{figure*}

In the case of $^{243}$U, as a 
result of projecting multidimensional fission K$_{0}$ = 9/2 paths
into the one-dimensional plot of the figure, the HFB-EFA 1F and 2F  curves 
appear as intersecting ones. However, in the multidimensional space of deformation 
parameters, there is a 9/2-path with a ridge connecting them. For  $^{243}$U 
as well as for all the other odd-mass nuclei 
studied in this work, we have 
neglected the contribution of such a path \cite{Rayner-UPRC-2014,Rayner-UEPJA-2014,Rayner-RaEPJA-2016}
to the collective action
Eq. (\ref{action}). This amounts to take 
the HFB-EFA 1F and 2F  curves as really intersecting in the computation of the 
spontaneous fission half-life.

In panel (c) of Fig.~\ref{peda-1}, we have plotted the  HFB-EFA and AV   
pairing energies. On the one hand, the proton pairing energies can hardly be 
distinguished. On the other hand the HFB-EFA and AV neutron pairing energies
display a similar trend with pronounced minima (maxima) around the 
$Q_{20}$ values corresponding to the ground state, the first and second 
fission isomers (the top of the inner and  second barriers). Nevertheless, the 
HFB-EFA neutron pairing energies tend to be smaller than the AV ones as a 
result of the 
quenching of pairing correlations via blocking. As a consequence 
of this quenching, there is an enhancement of the ATD and/or GCM masses with 
respect to the AV ones, as can be seen from panels (d) and (e), respectively.
We then conclude that the unpaired neutron leads to both the specialization energy effect 
and to the increase of the collective masses regardless of the ATD and/or GCM
scheme employed in their computation. Both effects go in the 
direction of increasing the collective action and, in turn, to larger 
spontaneous fission half-lives in odd-mass U and Pu nuclei as compared with their 
even-even neighbors (see, Sec.~\ref{syst-odd-mass-U-Pu}).

\begin{figure*}
\includegraphics[width=1.00\textwidth]{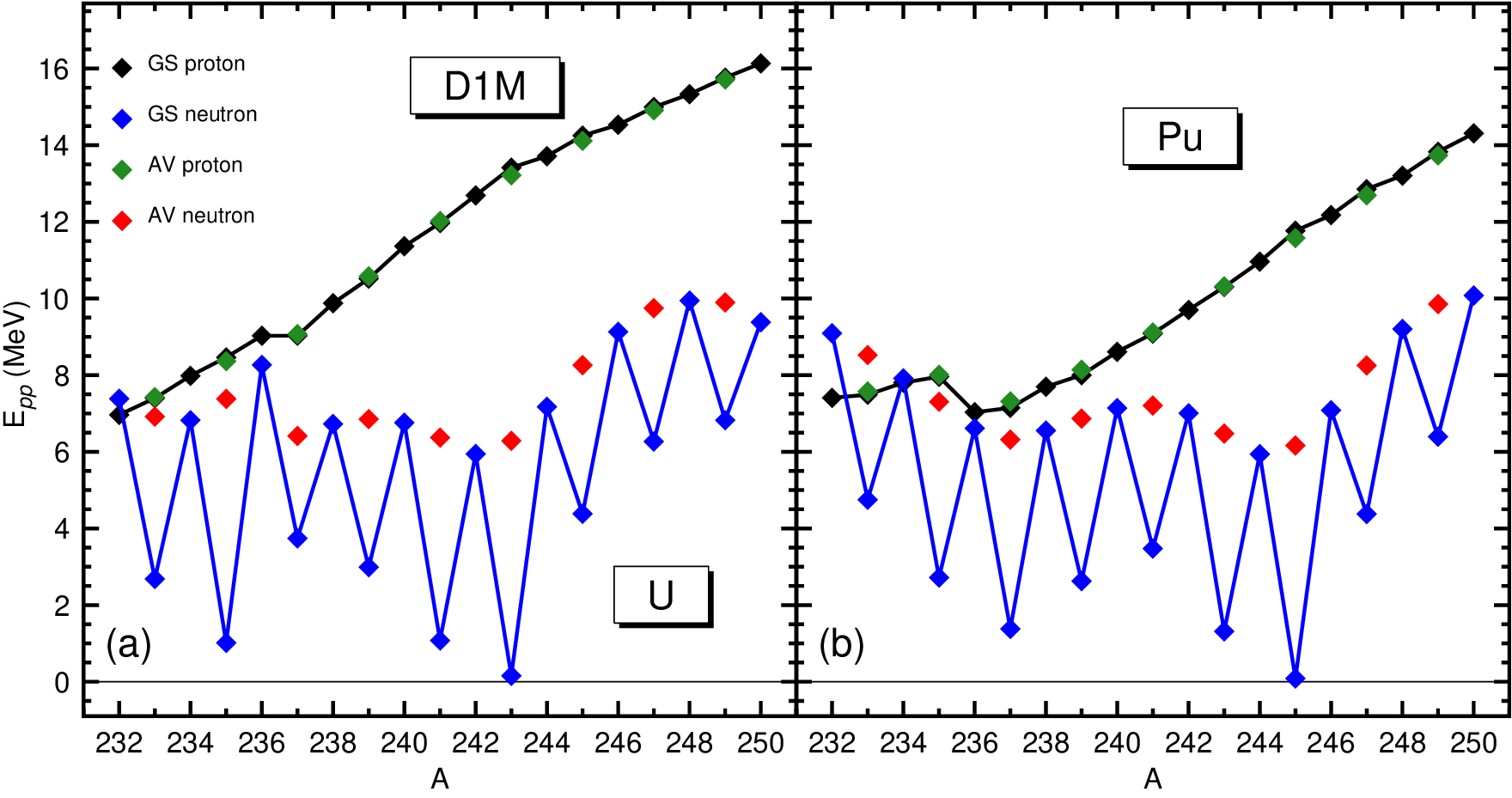}
\caption{(Color online) The proton and neutron pairing interaction energies 
E$_{pp}$ corresponding to the ground states (GS) of $^{232-250}$U [panel (a)]
and $^{232-250}$Pu [panel (b)] are plotted as functions of the mass number A.
Results  for  even-even U and Pu nuclei  
are taken from Refs.~\cite{Rayner-UPRC-2014} and \cite{Rayner-UEPJA-2014}. The
 "average" (AV) E$_{pp}$ values obtained for odd-mass nuclei are also 
included in the plots.
}
\label{PAIR} 
\end{figure*}

Furthermore, the ATD masses are larger than the GCM ones 
\cite{Rayner-UPRC-2014,Rayner-UEPJA-2014,Rayner-RaEPJA-2016,Rayner-No-odd}. In fact, such 
a difference is the reason why we have considered both kinds of
collective masses in the computation of the spontaneous fission half-lives
even, when there is a lack of theoretical justification for 
the use of the latter in the case of odd-mass nuclei. For example, in the case of  $^{243}$U and 
E$_{0}$ = 1.0 MeV, the ATD masses lead to  log$_{10}$ t$_{SF}$= 60.5530 s while the GCM ones lead to 
log$_{10}$ t$_{SF}$= 39.2109 s. On the other hand, increasing the value of E$_{0}$
provides a reduction in the t$_{SF}$ values. For example, for 
E$_{0}$ = 1.5 MeV, we have obtained  log$_{10}$ t$_{SF}$= 56.5407 s  and 
log$_{10}$ t$_{SF}$= 36.5494 s within the ATD and GCM schemes, respectively.

The density contour plots 
corresponding to the nucleus $^{243}$U at the quadrupole deformations 
$Q_{20}$ = 110 and 154 b are shown in panels (a), (b) and (c) of 
Fig.~\ref{con-den-243U}. For $Q_{20}$ = 154 b, two plots are shown corresponding 
to 1F and 2F solutions, respectively. The 2F solution in panel (c), corresponds 
to a spherical $^{132}$Sn fragment plus 
an oblate ($\beta_2$ = -0.21) and slightly octupole deformed 
($\beta_3$ = 0.02) $^{111}$Mo fragment. The shape of $^{111}$Mo minimizes 
a Coulomb repulsion energy of 186.50 MeV. Oblate deformed fragments 
have been obtained in previous studies 
\cite{Robledo-Giulliani,Rayner-UPRC-2014,Rayner-UEPJA-2014,Rayner-RaEPJA-2016} as well as 
 by fissioning other odd-mass U and Pu nuclei 
as we will see later on in this paper
(see, Sec.~\ref{syst-odd-mass-U-Pu}). The previous results, illustrate the key role 
played by the shell effects associated with the proton Z=50 and neutron N =82 
magic numbers \cite{Nenoff-2007,Ter-1996,Piessens-1993} in determining 
the charge and mass of the fission fragments. On the one hand, this could
be expected within the framework  of Ritz-variational approaches 
\cite{Rayner-UPRC-2014,Rayner-UEPJA-2014,Rayner-RaEPJA-2016} and, in particular,   
the HFB-EFA. On the other hand, the  available data \cite{Pu-mass-fragments-exp-1,Pu-mass-fragments-exp-2} indicate 
that, for the considered region,  the heavy fragment mass number is close to A=140 instead 
of A=132. In our calculations the properties of the 
fragments are determined by 2F solutions of the HFB-EFA equations at the largest Q$_{20}$ 
values (see, Fig.~\ref{peda-1}). Nevertheless, such 2F solutions 
are not necessarily the ones corresponding to scission products. A more realistic
approximation to the mass and charge of the fission fragments has to include 
dynamical effects around the loosely defined scission configuration
\cite{Goutte-dynamical-distribution,Chasman-breaking}. 
Moreover, our calculations do not account for the broad mass distribution
observed experimentally. Therefore, the values obtained should be taken only as an 
approximation to the  peaks in that distribution.
 
\begin{figure*}
\includegraphics[width=1.00\textwidth]{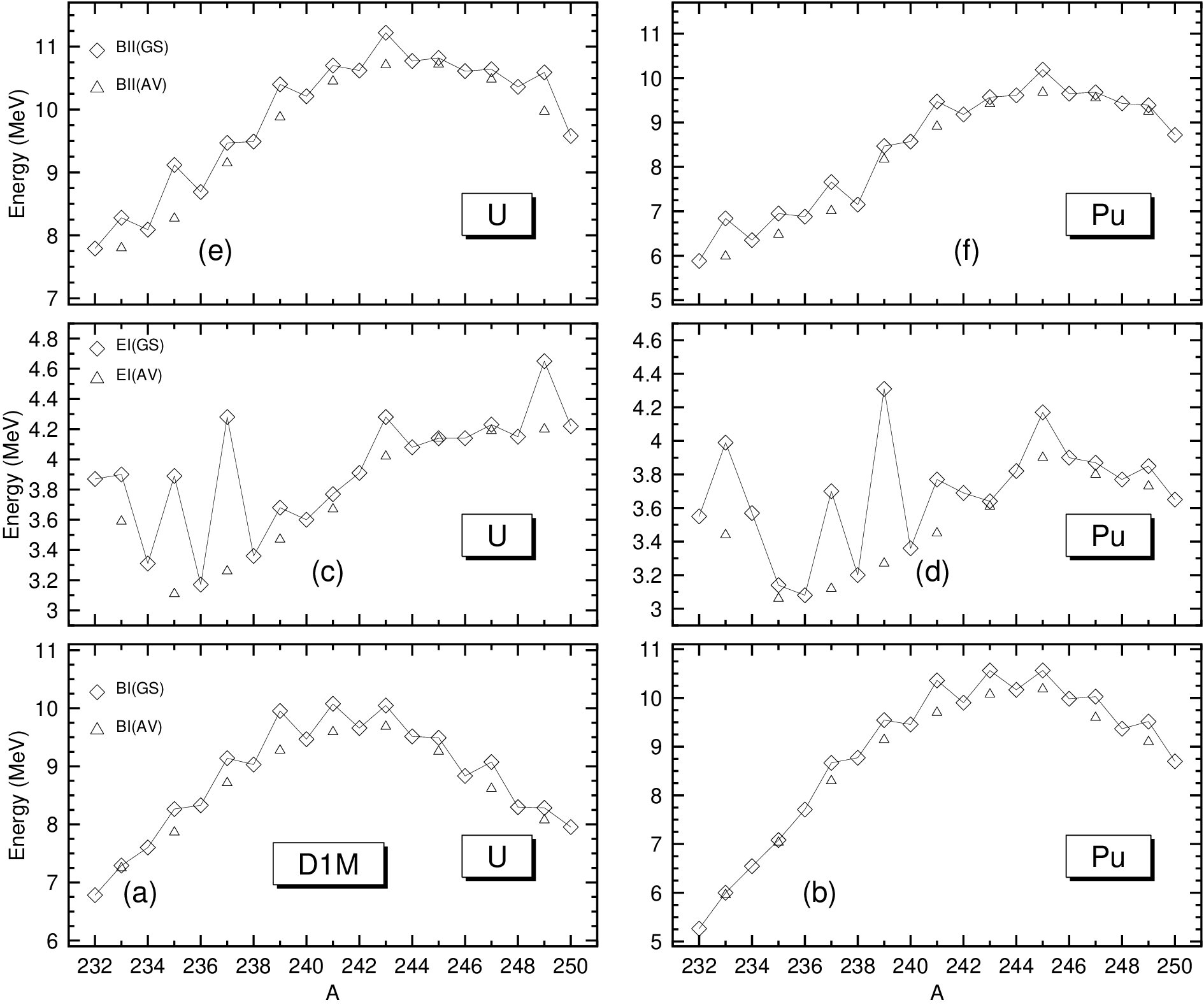}
\caption{The inner barrier height B$_{I}$ [panels (a) and (b)], excitation energy of the first 
fission isomer  E$_{I}$ [panels (c) and (d)] and the second barrier height B$_{II}$
[panels (e) and (f)] corresponding to
the ground-state (GS) fission paths in odd mass U and Pu isotopes with A=233,\ldots,249
are plotted as functions of the mass number A. 
Results  for  even-even U and Pu nuclei  
are taken from Refs.~\cite{Rayner-UPRC-2014} and \cite{Rayner-UEPJA-2014}. The "average"
(AV) B$_{I}$, E$_{I}$ and B$_{II}$ values obtained for odd-mass U and Pu nuclei 
are also included in the plot.
}
\label{BI_BII_U_PU} 
\end{figure*}

\subsection{Systematic of the fission paths, spontaneous fission half-lives and 
fragment mass and charge in odd-mass U and Pu nuclei}
\label{syst-odd-mass-U-Pu}

In Fig.~\ref{bariers-U}, we have summarized the ground state fission paths obtained 
for the nuclei $^{233,235,237,239}$U [panel (a)] and 
$^{241,243,245,247,249}$U [panel (b)]. The  fission paths for the 
isotopes $^{233,235,237,239}$Pu and 
$^{241,243,245,247,249}$Pu are shown in panels (a) and (b) of Fig.~\ref{bariers-Pu}. The corresponding K$_{0}$ values are 
also given in the plots.  
The AV fission paths obtained for those
odd-mass nuclei are  depicted with dashed lines. 
Results  for  even-even U and Pu nuclei  
are taken from Refs.~\cite{Rayner-UPRC-2014} and \cite{Rayner-UEPJA-2014}.
The fission 
paths for odd-mass U and Pu nuclei exhibit a  structural evolution 
similar to the  one obtained for the corresponding even-even counterparts. The ground 
state is located around Q$_{20}$ = 14 b while first fission isomers are found 
in the range of
quadrupole moments 36 $\le$ Q$_{20}$ $\le$ 44 b. Those first 
isomers are separated from the ground 
state by the corresponding inner barriers the tops of which correspond to
22 $\le$ Q$_{20}$ $\le$ 30 b. The heights of those inner  barriers can be 
reduced, by a few MeV, due to triaxiality \cite{Rayner-UPRC-2014}. Moreover, in the case
of odd-mass nuclei the polarization effects associated with the unpaired nucleon might 
also lead to triaxial solutions. In our calculations, we have kept axial symmetry 
as a self-consistent symmetry along the whole fission path 
in order to reduce the   computational effort. On the other hand, the 
already mentioned lowering of the inner barriers, by the $\gamma$ degree of 
freedom, comes together with an increase of the collective inertia 
\cite{Baran-1981,Bender-1998}
that tends to compensate
in the final value of the action Eq. (\ref{action}). Therefore, the role 
of the $\gamma$ deformation parameter in the spontaneous fission half-lives 
is very limited \cite{Baran-1981}. On the other hand, the tops of the 
second barriers correspond to
52 $\le$ Q$_{20}$ $\le$ 70 b.

From the figures, one also realizes 
that second fission isomers appear for several even-even and odd-mass U and 
Pu nuclei indicating that the corresponding shell effects are systematically
present in our Gogny-D1M calculations for this region of the nuclear chart
\cite{Rayner-UPRC-2014,Rayner-UEPJA-2014,Rayner-RaEPJA-2016}. 
Those 
parity-breaking 
second isomers become apparent around 
$^{239}$U and $^{243}$Pu and their quadrupole moments 
lie within the range 
84 $\le$ Q$_{20}$ $\le$ 96 b. With increasing neutron number, outer 
barriers also emerge along the 1F paths with their tops 
corresponding to quadrupole deformations Q$_{20}$ $\ge$ 110 b. Furthermore, the 
comparison between the HFB-EFA  ground state and AV (1F and 2F) paths  reveals the specialization energy
effects in the case of  odd-mass systems.

\begin{figure*}
\includegraphics[width=1.0\textwidth]{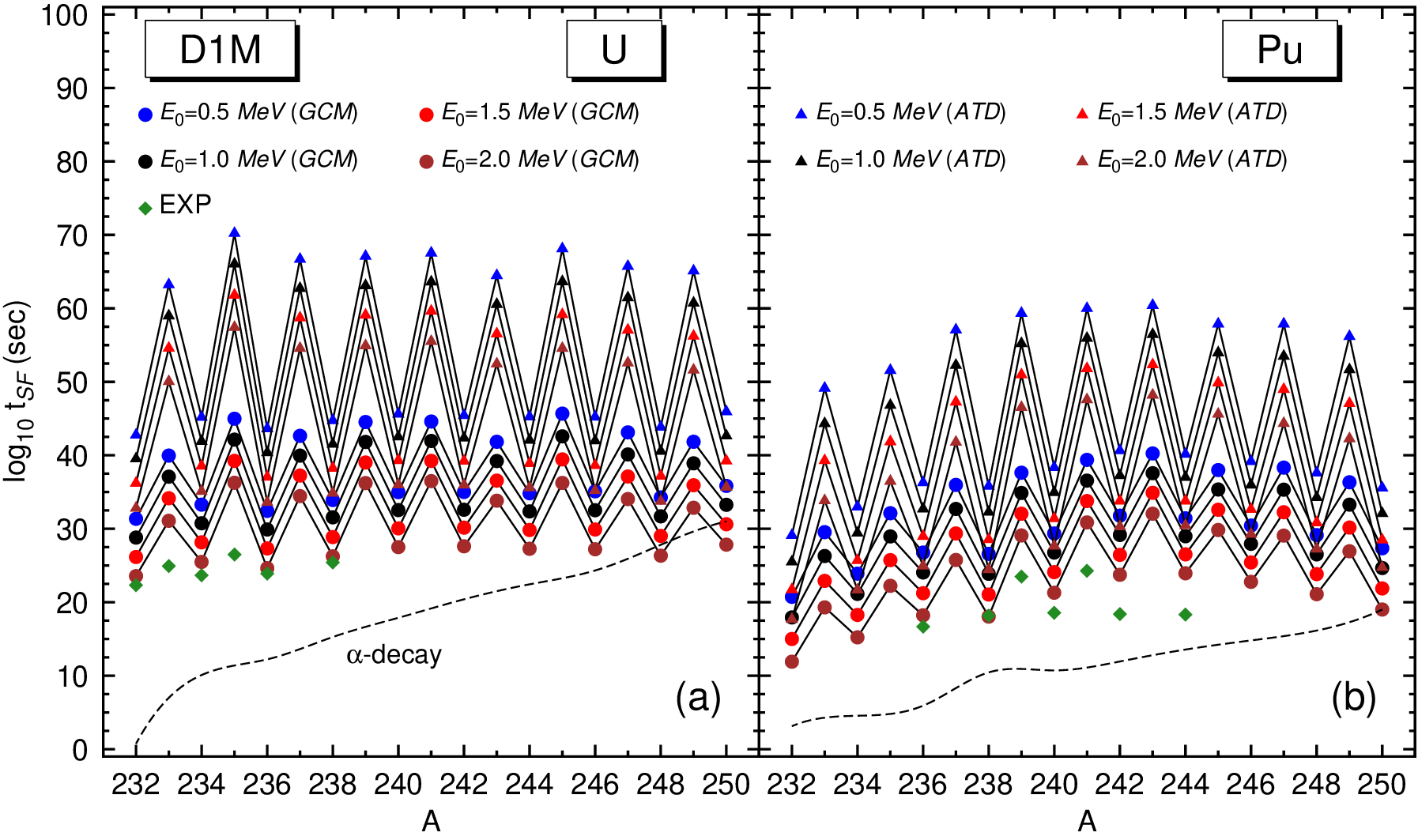}
\caption{(Color online) The spontaneous fission half-lives, predicted within the  
GCM  and ATD  schemes, for the nuclei $^{232-250}$U 
[panel (a)] and $^{232-250}$Pu  [panel (b)]
are depicted 
as functions of the mass number A. For each nucleus, calculations have been 
carried out with E$_{0}$ = 0.5, 1.0, 1.5 and 2.0 MeV. Results  for  even-even U and Pu nuclei  
are taken from Refs.~\cite{Rayner-UPRC-2014} and \cite{Rayner-UEPJA-2014}. The experimental
t$_{SF}$ values are taken from Ref.~\cite{Holden-paper}. In addition, $\alpha$-decay half-lives 
are plotted with short dashed lines. For more details, see the main text.
}
\label{tsf-U} 
\end{figure*}

From the experimental point of view the ground states
of the nuclei $^{233,235,237,239}$U correspond to 
5/2$^{+}$, 7/2$^{-}$, 1/2$^{+}$
and 5/2$^{+}$ configurations, respectively. On the other hand, the 
nuclei $^{235,237,239,241, 243, 245,247}$Pu
have 5/2$^{+}$, 7/2$^{-}$, 1/2$^{+}$, 5/2$^{+}$, 7/2$^{+}$, 9/2$^{-}$ 
and 1/2$^{+}$
ground states \cite{EXP-Spins-Gs-U-Pu}. Our calculations, agree well 
with the available  data for $^{233,239}$U and
$^{235,241,243,245,247}$Pu. On the other hand, given the discrepancy 
between our predictions and the experimental K$_{0}$ values, in the case of  
 $^{235}$U  and $^{237}$U ($^{237}$Pu and $^{239}$Pu)
 we  have also  explored 
 the K = 7/2 and 1/2  paths. They are plotted in 
 Fig.~\ref{bariers-235U237U237Pu239Pu}  as functions of
 the quadrupole moment Q$_{20}$. Both their structure and the 
 corresponding specialization energy effects are similar 
 to the ones for the ground state fission paths in 
 Figs.~\ref{bariers-U} and \ref{bariers-Pu}.

The proton and neutron ground state (GS) pairing interaction energies 
E$_{pp}$ \cite{rs} 
are plotted in panels (a) and (b) of 
Fig.~\ref{PAIR} as functions of the mass number A. The AV
E$_{pp}$ values obtained for  odd-mass nuclei are also
included in the plots. The HFB-EFA and AV proton pairing energies 
are rather similar and exhibit a sharp increase with increasing mass number 
A. On the other hand, the quenching of the HFB-EFA neutron pairing with respect to 
the AV one, via blocking, is rather pronounced. As a measure of how effective is 
the quenching of the neutron pairing correlations 
in the ground states of the   odd-mass systems,
one can take the ratio r=$\langle \Delta \hat{N}^{2} \rangle_{GS}$/$\langle \Delta 
\hat{N}^{2} \rangle_{AV}$ \cite{Rayner-No-odd}. We have obtained the values 
r=0.360, 0.367, 0.579, 0.667, 0.631, 0.103, 0.661, 0.839 and 
0.814 for the U isotopes with A=233,\ldots,249 while for the Pu isotopes
with the same mass numbers the corresponding values are
r=0.817, 0.457, 0.380, 0.631, 0.717, 0.332, 0.07, 0.653 and 0.843.
For the absolute minima of the K = 7/2 and 1/2   paths (Fig.~\ref{bariers-235U237U237Pu239Pu}) in 
$^{235}$U  and $^{237}$U ($^{237}$Pu and $^{239}$Pu)
we have obtained the ratios 
r = 0.70 and 0.78 (0.65 and 0.77).
From the previous results we conclude that, at least for some 
of the studied odd-mass U and/or Pu nuclei, a more realistic treatment of pairing (including 
the role of 
dynamical  fluctuations and their coupling to the relevant deformation 
parameters \cite{Action-Rayner}) is required to describe the fission process. Work 
along these lines is in progress and will be
reported elsewhere.

\begin{table*}
\caption{The values of log$_{10}$ t$_{SF}$ (in s) obtained for the K = 7/2 path in $^{235}$U ($^{237}$Pu)  and 
the K = 1/2 path in 
$^{237}$U  ($^{239}$Pu)
are given as functions of the parameter E$_{0}$ (in MeV). For details, see the main text.
} 
\begin{tabular}{ccccccc}
\hline
Nucleus     & K       & Scheme          & E$_{0}$=0.5 MeV   & E$_{0}$ = 1.0 MeV & E$_{0}$ = 1.5 MeV  & E$_{0}$ = 2.0 MeV  \\
\\
\hline
$^{235}$U   & 7/2     &  GCM            &  40.2335          &  37.4653          &   34.6713          &    31.7890           \\ 
\\ 
$^{235}$U   & 7/2     &  ATD            &  65.6648          &  61.5757          &   57.4544          &    53.2159           \\   
\\ 
$^{237}$U   & 1/2     &  GCM            &  44.2782          &  41.2471          &   38.1644          &    34.9754           \\
\\ 
$^{237}$U   & 1/2     &  ATD            &  68.1143          &  63.7326          &   59.2679          &    54.6358            \\
\hline
\hline
$^{237}$Pu  & 7/2     &  GCM	        &  34.3406	    &  31.4090	        &   28.3981	     &	 25.2630 	  \\
\\
$^{237}$Pu  & 7/2     &  ATD	        &  56.4852	    &  52.1555	        &   47.7064	     &	 43.0442 	 \\
\\
$^{239}$Pu  & 1/2     &  GCM	        &  36.2572	    &  33.0818	        &   29.7459	     &	 26.1170 	  \\
\\
$^{239}$Pu  & 1/2     &  ATD	        &  57.2207	    &  52.5042	        &   47.4956	     &   42.0561	   \\
\hline
\end{tabular}
\label{other-tsf-233-241U-233-241Pu}
\end{table*}

The inner barrier heights B$_{I}$ [panels (a) and (b)], excitation energies of the first 
fission isomers  E$_{I}$ [panels (c) and (d)] and the second barrier heights B$_{II}$
[panels (e) and (f)] are depicted in 
Fig.~\ref{BI_BII_U_PU}. The 
AV B$_{I}$, E$_{I}$ and B$_{II}$ values 
obtained for odd-mass nuclei
are also included in the plots. With few exceptions, the HFB-EFA
values exhibit 
odd-even as well as the
specialization energy effects already discussed in the 
case of $^{243}$U (Sec.~\ref{example-methodology}).
Similar trends are obtained for the excitation energies 
of the second fission isomers and the heights of the outer
barriers  in the case of heavier nuclei.
Note that, as already discussed above, the heights of the inner 
barriers might be reduced, by a few MeV, due to triaxiality
\cite{Rayner-UPRC-2014}. In our calculations the largest 
B$_{I}$, E$_{I}$ and B$_{II}$ values 
of 10.07 (10.56), 4.65 (4.31) and 11.22 (10.19) MeV
are obtained for $^{241}$U ($^{243}$Pu), $^{249}$U ($^{239}$Pu)
and $^{243}$U ($^{245}$Pu), respectively. 
Furthermore, we have obtained the B$_{I}$, E$_{I}$ and  B$_{II}$ values 
of 7.88, 4.84 and  8.62 MeV (8.40, 4.49 and 7.64 MeV) for   
the K = 7/2 path in $^{235}$U ($^{237}$Pu)  while 
the corresponding values for the K = 1/2 path in 
$^{237}$U  ($^{239}$Pu) are 8.91, 3.38 and 9.20 MeV
(9.40, 3.32 and 8.29 MeV). These values, should be 
compared with the ones shown in the figure, i.e., 
8.27, 3.89 and 9.12 MeV (8.66, 3.70 and 7.76 MeV) for the ground state fission 
path of $^{235}$U ($^{237}$Pu) and 9.14, 4.28 and 9.47 MeV
(9.54, 4.31 and 8.47 MeV)
for the ground state path of $^{237}$U ($^{239}$Pu).

For all the 
studied odd-mass nuclei, we have also obtained a pronounced enhancement 
of the ATD and/or GCM collective masses, with respect to the AV
ones, which results
from the quenching of the neutron pairing 
correlations by the unpaired neutron (see, Figs.~\ref{peda-1} 
and \ref{PAIR}). Both 
the specialization energy effects described above  and 
the enhancement of the collective masses, 
are the main factors 
leading, as we will see later on, to larger spontaneous fission half-lives in
the odd mass U and Pu isotopes considered, as compared with their 
even-even neighbors \cite{Rayner-No-odd}.

In Fig.~\ref{tsf-U}, we have depicted 
the spontaneous fission half-lives, predicted within the  
GCM  and ATD  schemes, for U 
[panel (a)] and Pu  [panel (b)] nuclei
as functions of the mass number A. For each nucleus, calculations have been 
carried out with E$_{0}$ = 0.5, 1.0, 1.5 and 2.0 MeV. 
The experimental values shown 
in the figure for $^{232}$U and $^{233}$U correspond to lower bounds. 
No experimental t$_{SF}$ value is available for $^{237}$U
\cite{Holden-paper}. In the case of $^{241}$Pu the value shown 
in the figure corresponds to an upper bound while
no experimental data are available for
$^{232,233,234,235,237,243}$Pu  \cite{Holden-paper}. The
data reveal an increase in the spontaneous 
fission half-lives of the odd-mass 
nuclei as compared with their even-even 
neighbors. Such a trend is also observed in our 
calculations. Increasing the parameter E$_{0}$ leads 
to a decrease in the predicted t$_{SF}$ values as well as to an improvement 
of the agreement with the experiment. That larger E$_{0}$ values are required 
for a better comparison with the experiment is, a consequence of the fact
that the Gogny-D1M EDF provides wide 1F curves with a gentle decline 
\cite{Rayner-UPRC-2014}. In Sec.~\ref{pairstrength-odd-mass-U-Pu}, we will
also discuss the impact of pairing correlations on the predicted 
t$_{SF}$ values.

The ATD spontaneous fission half-lives are larger 
than the GCM ones, with the difference being more 
pronounced for odd-mass nuclei. For example, for 
$^{247}$U and $^{247}$Pu with E$_{0}$ = 1.0 MeV we have 
obtained 
log$_{10}$ t$_{SF}$= 61.4421 s and log$_{10}$ t$_{SF}$= 53.5035 s
within the ATD scheme while the corresponding GCM values 
are $\log_{10}$ t$_{SF}$= 40.1248 s
and $\log_{10}$ t$_{SF}$= 35.3381 s, respectively
(see also, table~\ref{other-tsf-233-241U-233-241Pu}). The previous results already reveal  
the strong variance of the predicted t$_{SF}$ values 
with respect to the details involved in their 
computation (see also, Sec.~\ref{pairstrength-odd-mass-U-Pu}). 
However, regardless of the scheme and/or E$_{0}$ value 
employed, the same trend emerges from our calculations, i.e., the 
fission lifetimes exhibit a pronounced odd-even staggering. 
For a given odd-mass nucleus, the 
amplitude of the staggering can be defined as 
\begin{eqnarray}
\delta_{st} = \log_{10} t_{SF}(odd) - \log_{10} t_{SF}(ee,av) 
\end{eqnarray}
where $t_{SF}(odd)$ represents its spontaneous fission 
half-life while $t_{SF}(ee,av)$ is the average value for the
two even-even neighbors. We have studied such a quantity 
and found that it depends strongly on both the 
neutron number and the type of collective 
mass employed, with the ATD values $\delta_{st}^{ATD}$ being larger than the GCM 
$\delta_{st}^{GCM}$ ones. For example  for the nuclei $^{233}$U ($^{233}$Pu)
and $^{241}$U ($^{241}$Pu) we have obtained, with E$_{0}$ = 1.0 MeV, the ATD values 
$\delta_{st}^{ATD}$ = 17.3711 s (15.2123 s) and $\delta_{st}^{ATD}$ = 21.1714 s (18.9971 s)
while the GCM staggering for those nuclei are 
$\delta_{st}^{GCM}$ = 6.6337 s (5.4624 s) and 9.3843 s (7.6828 s), respectively. On the other hand, the 
amplitude of the staggering is rather insensitive to  E$_{0}$. Note, that similar features 
to the ones already mentioned emerge if we consider (see table~\ref{other-tsf-233-241U-233-241Pu}) the K = 7/2 path in $^{235}$U ($^{237}$Pu)  and 
the K = 1/2 path in 
$^{237}$U  ($^{239}$Pu).

In addition, in Fig.~\ref{tsf-U} we have  plotted 
the $\alpha$-decay half-lives 
computed with a parametrization  \cite{TDong2005} of the 
Viola-Seaborg formula \cite{Viola-Seaborg}. We have used 
the binding energies obtained for U, Pu and Th nuclei.
We conclude that, though $\alpha$-decay is the dominant 
decay channel for most of the studied nuclei, the steady 
increase in the t$_{\alpha}$ values as functions of the mass number A, leads
 to a pronounced competition with 
spontaneous fission around A = 248-250.
This agrees well with previous fission 
calculations for Ra \cite{Rayner-RaEPJA-2016}, U \cite{Rayner-UPRC-2014} 
and Pu \cite{Rayner-UEPJA-2014}
nuclei which suggest that as we move to the very neutron-rich sectors 
of the corresponding isotopic chains, fission turns out to be faster 
than $\alpha$-decay.

\begin{figure*}
\includegraphics[width=1.0\textwidth]{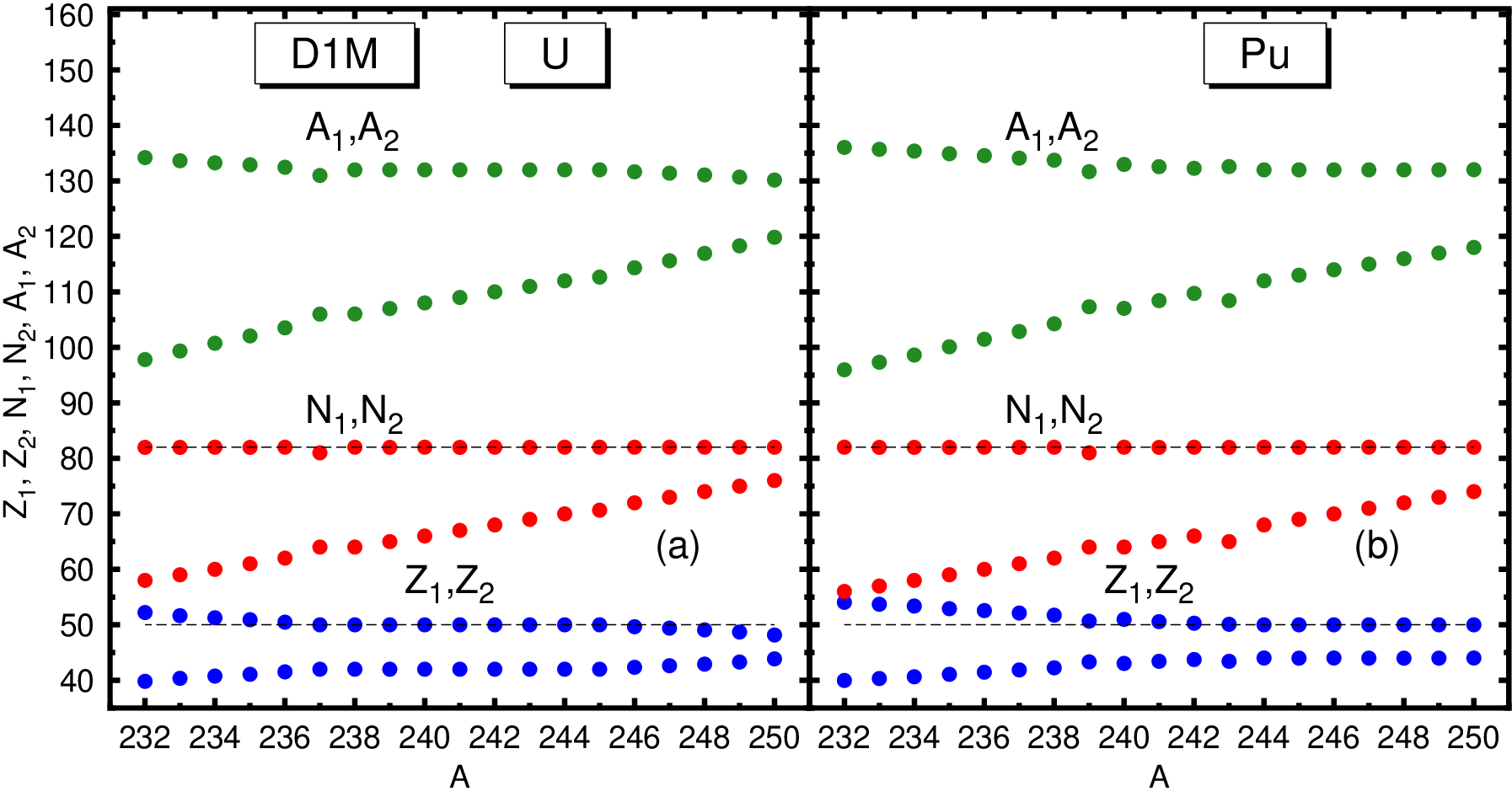}
\caption{(Color online) The proton ($Z_{1},Z_{2}$), neutron ($N_{1},N_{2}$) and mass ($A_{1},A_{2}$)
numbers of the two  fragments resulting from the fission of $^{232-250}$U [panel (a)]
and $^{232-250}$Pu [panel (b)] are shown as functions of 
the mass number A in the  parent nucleus. Results  for  even-even U and Pu nuclei  
are taken from Refs.~\cite{Rayner-UPRC-2014} and \cite{Rayner-UEPJA-2014}. The magic 
proton $Z=50$ and  neutron $N=82$ numbers
are highlighted with dashed horizontal lines to guide the eye.
}
\label{fragments-U-Pu} 
\end{figure*}

The proton (Z$_{1}$, Z$_{2}$), neutron
(N$_{1}$, N$_{2}$) and mass (A$_{1}$, A$_{2}$) numbers 
of the 
fission fragments are plotted, as functions of the mass number A
in the parent nucleus, in Fig.~\ref{fragments-U-Pu}. The key role 
played by the proton Z=50 and neutron N=82 magic numbers in the masses and charges 
of the predicted fission fragments is apparent from the figure. However, 
the properties
of those fragments are determined by 
Ritz-variational 
solutions of the HFB-EFA equations 
along the 2F curves (see, Figs.~\ref{peda-1}, \ref{bariers-U} and 
\ref{bariers-Pu}) at the largest quadrupole moments. Therefore, caution
should be taken when comparing with available experimental
data 
for this region of the nuclear chart
(see, for example, \cite{Pu-mass-fragments-exp-1,Pu-mass-fragments-exp-2,Schmidt-fragments}).

\begin{figure*}
\includegraphics[width=0.3\textwidth]{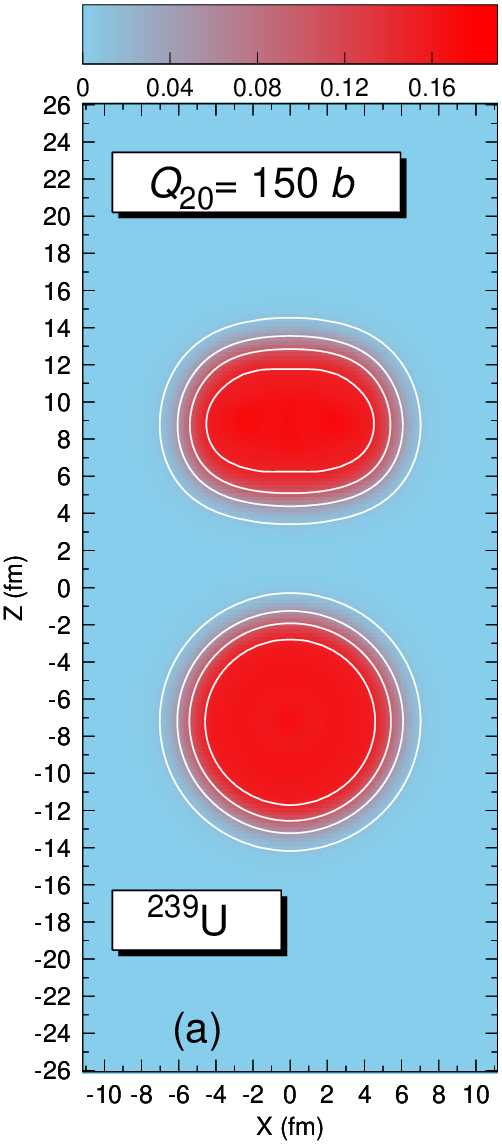}
\includegraphics[width=0.3\textwidth]{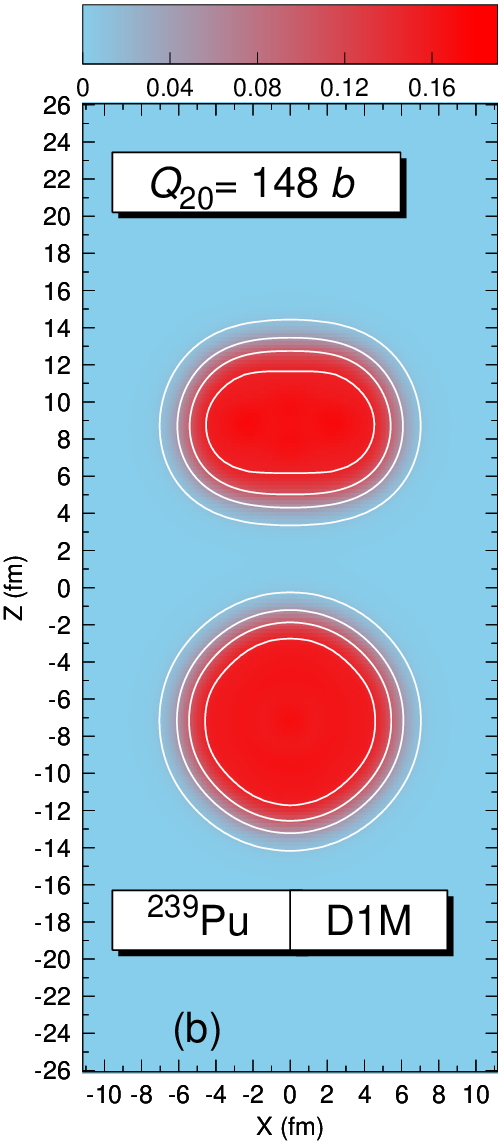}
\includegraphics[width=0.3\textwidth]{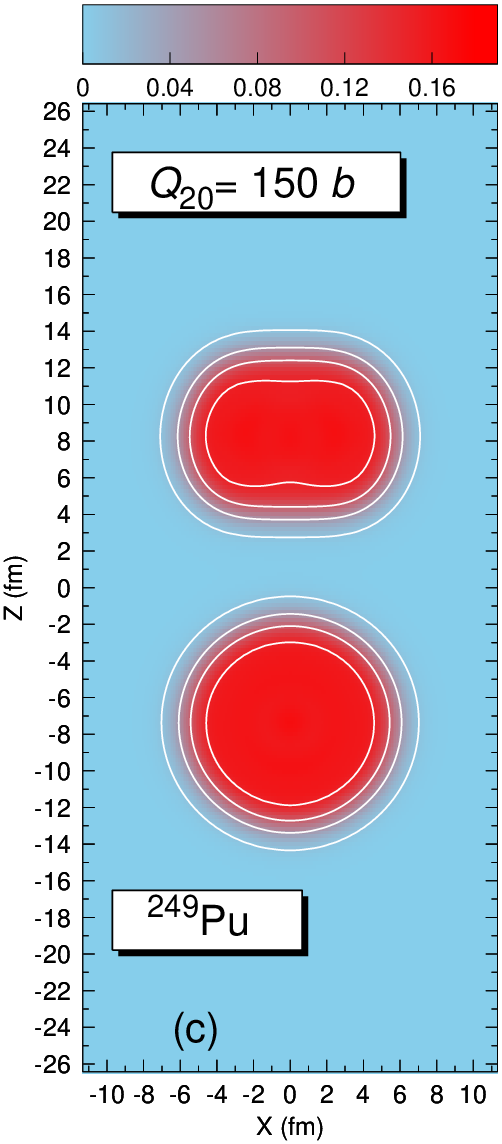}
\caption{(Color online) Density contour plots for the 
nuclei $^{239}$U [panel (a)], $^{239}$Pu [panel (b)]
and $^{249}$Pu [panel (c)]. The density profiles correspond 
to 2F solutions at the quadrupole deformations 
Q$_{20}$=150, 148 and 150 b.  
Densities are in units of fm$^{-3}$ and contour lines are drawn at 0.01, 0.05, 0.10 and 
0.15 fm$^{-3}$.
For more details, see the main text.  
}
\label{con-den-others} 
\end{figure*}

We have also studied the shapes of the fission fragments.
As illustrative examples, we have plotted
in Fig.~\ref{con-den-others}  the 
density contour plots for  
 $^{239}$U [panel (a)], $^{239}$Pu [panel (b)]
and $^{249}$Pu [panel (c)] at the 
quadrupole deformations 
Q$_{20}$=150, 148 and 150 b, respectively. For both $^{239}$U and 
$^{249}$Pu, a spherical $^{132}$Sn heavier fragment is predicted while 
the lighter fragments correspond to oblate and slightly 
octupole deformed  $^{107}$Mo ($\beta_2$ = -0.23, $\beta_3$ = 0.02)
 and $^{117}$Ru ($\beta_2$ = -0.19, $\beta_3$ = 0.02)
 nuclei, respectively. In the case of $^{239}$Pu our calculations 
 predict an almost spherical  heavier fragment with Z = 50.67 and N = 80.99
 while the deformed ($\beta_2$ = -0.23, $\beta_3$ = 0.02)
 lighter fragment has Z = 43.33 and N = 64. The predicted oblate 
 fragments in $^{239}$U, $^{239}$Pu 
and $^{249}$Pu minimize  large Coulomb repulsion energies 
 of 187.00, 196.79 and 200.66 MeV. The appearance of oblate fragments 
 in our calculations
 for even-even 
 \cite{Rayner-UPRC-2014,Rayner-UEPJA-2014}
 and odd-mass U and Pu nuclei,  as well as for other systems
 in this region of the nuclear chart \cite{Rayner-RaEPJA-2016}, deserves 
 further attention as fission fragments are usually 
 assumed to have  prolate shapes \cite{Moller-1,Moller-2}. 

\subsection{Varying pairing strengths  in odd-mass U and Pu nuclei}
\label{pairstrength-odd-mass-U-Pu}

In this section, we discuss
the impact of pairing correlations on the predicted spontaneous fission half-lives.
We have carried out  calculations, along the lines discussed in 
Sec.~\ref{Theory}, but with a modified Gogny-D1M EDF in which a factor $\eta$ has been 
introduced in front of the pairing fields \cite{rs}. For simplicity, we have considered the same 
$\eta$ = 1.05 and 1.10  factor for both protons and neutrons \cite{Rayner-UPRC-2014}. Let us mention, that the main reason to consider
modified  strengths is that, as we have seen in 
Secs.~\ref{example-methodology} and \ref{syst-odd-mass-U-Pu}, pairing
correlations are key ingredients in the computation of the 
collective masses as well as the zero-point rotational and 
vibrational quantum 
quantum corrections 
\cite{proportional-1,proportional-2,Rayner-RaEPJA-2016,Rayner-UPRC-2014,Rayner-UEPJA-2014}.

\begin{figure*}
\includegraphics[width=0.46\textwidth]{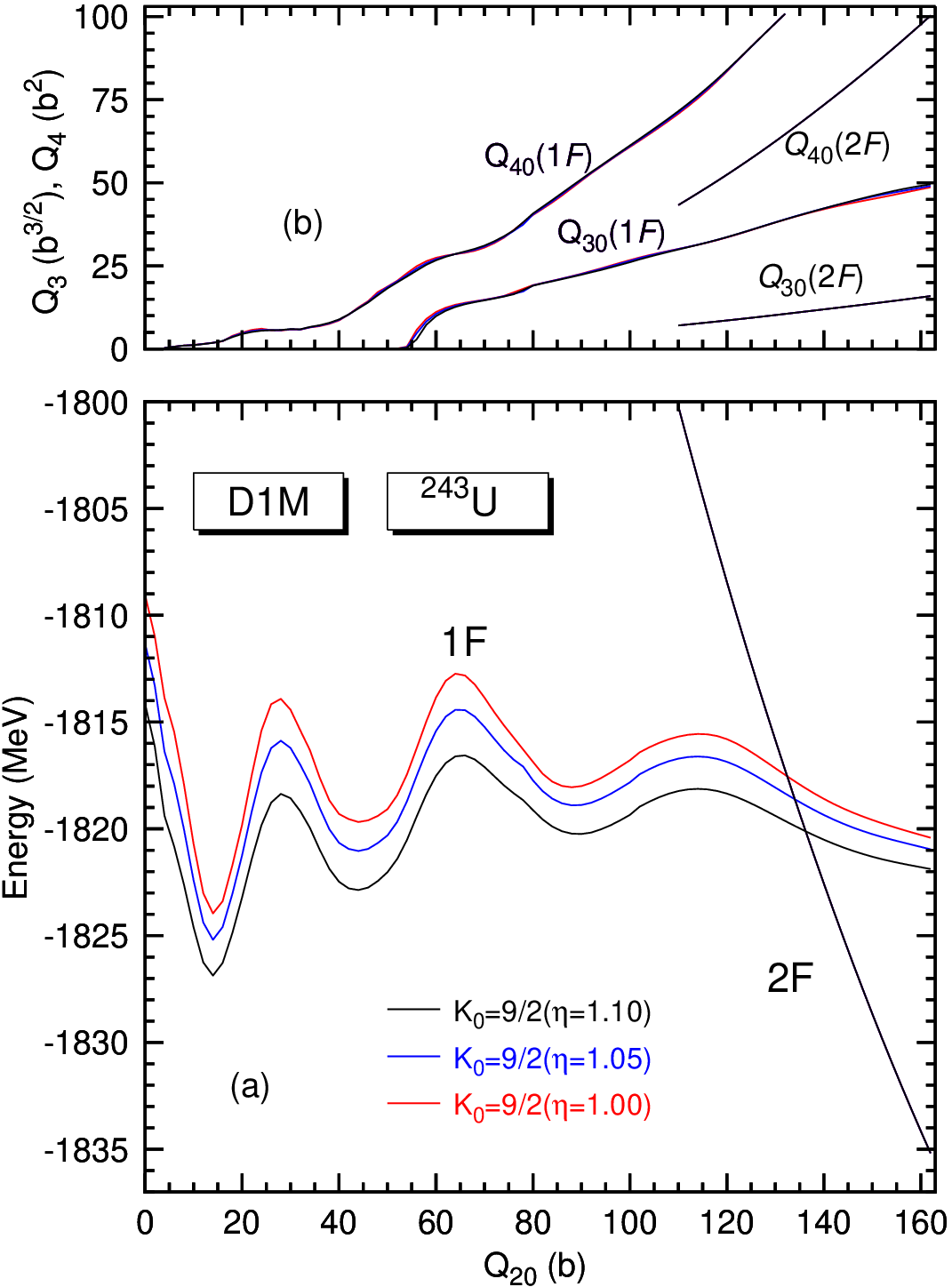}
\includegraphics[width=0.525\textwidth]{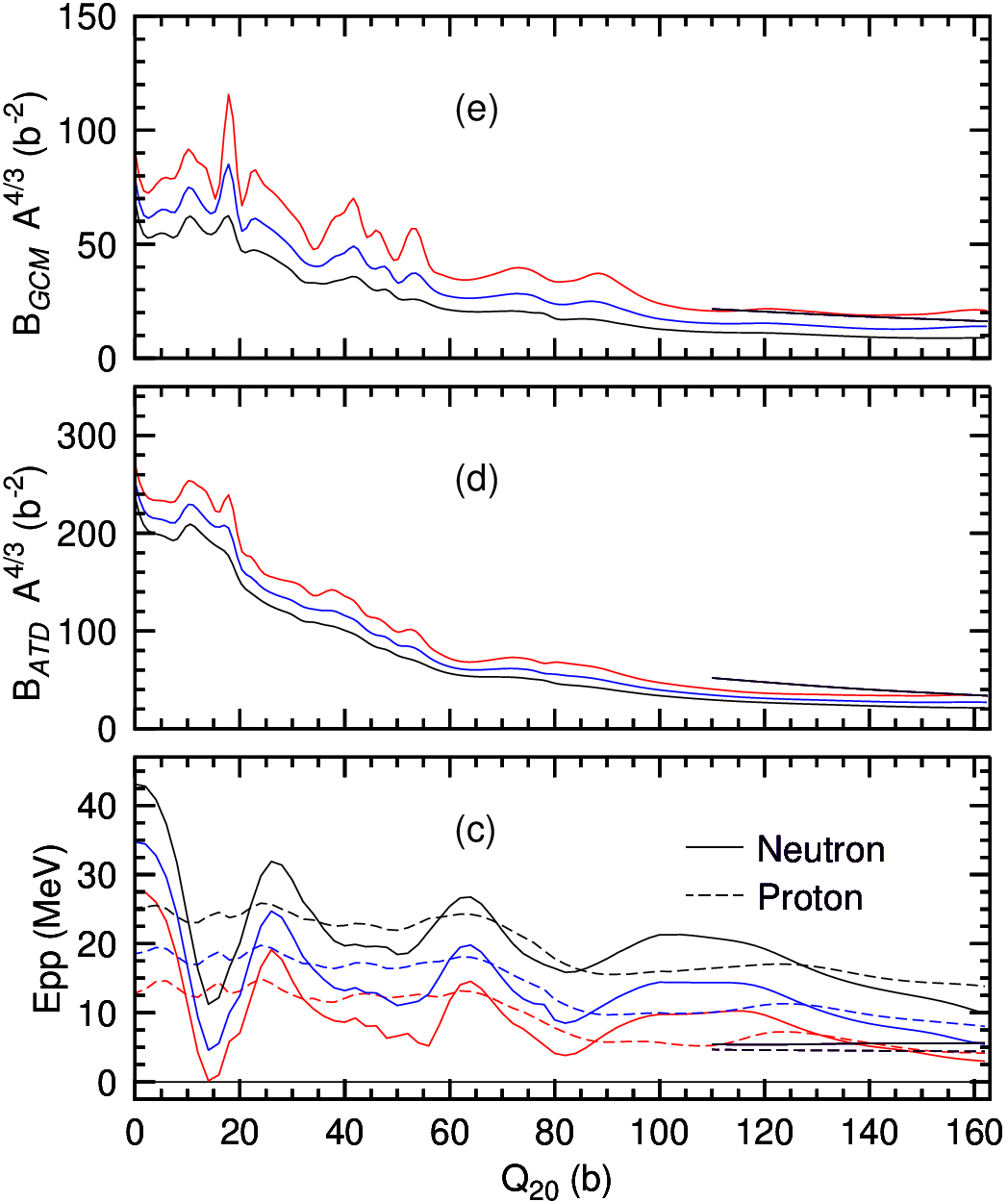}
\caption{(Color online) The K$_{0}$ = 9/2 HFB-EFA plus the zero point rotational energies
obtained with the normal ($\eta$=1.00) and modified ($\eta$=1.05 and $\eta$=1.10) Gogny-D1M EDF
 are plotted in panel 
(a) as functions of the quadrupole moment Q$_{20}$ for the nucleus $^{243}$U. 
For each $\eta$ value, both the one (1F) and two-fragment (2F) solutions are included 
in the plot. The octupole Q$_{30}$
and hexadecapole Q$_{40}$ moments
corresponding to the 1F and 2F solutions 
 are shown in panel (b). The pairing interaction energies are depicted 
in panel (c) for protons (dashed lines) and neutrons (full lines). The collective ATD and GCM masses are plotted
in panels (d) and (e), respectively. For more details, see the main text.  
}
\label{peda-1-eta-pair} 
\end{figure*}

The K$_{0}$ = 9/2 HFB-EFA plus the zero point rotational energies
obtained with the normal ($\eta$=1.00) and modified ($\eta$=1.05 and $\eta$=1.10) Gogny-D1M EDF
 are plotted in panel 
(a) of Fig.~\ref{peda-1-eta-pair}  as functions of the quadrupole moment Q$_{20}$ for the nucleus 
$^{243}$U, taken as an illustrative example. Similar 
calculations have been carried out for all
the nuclei studied in this paper. For each $\eta$ value, both the 1F and 2F solutions are included 
in the plot. 
The octupole Q$_{30}$
and hexadecapole Q$_{40}$ moments
corresponding to the 1F and 2F solutions 
 are shown in panel (b) of the figure. The 1F and 2F curves, for $^{243}$U and 
 all the studied odd-mass nuclei, exhibit rather similar shapes for different 
 $\eta$-values. Note, however, that the energy shifts obtained for $\eta$ = 1.05 and/or
 $\eta$ = 1.10 depend on the quadrupole moment. For example, in the case of  
 $^{243}$U, the energy gain for the K$_{0}^{\pi}$ = 9/2$^{-}$ ground state 
 is 1.23 (2.91) MeV  while the heights of the inner and second barriers 
 are reduced  by 0.73 (1.55) and 0.46 (0.92) MeV for $\eta$ = 1.05 
 ($\eta$ = 1.10) when compared wit the values obtained with the normal 
Gogny-D1M EDF. On the other hand, the octupole and hexadecapole moments 
of both the 1F and 2F solutions, can hardly be distinguished for 
different $\eta$-values. In panel (c) of the figure, we have depicted the pairing interaction energies 
for protons (dashed lines) and neutrons (full lines). We observe the same
trend though the pairing energies increase with increasing $\eta$-values.

The collective ATD and GCM masses are plotted
in panels (d) and (e) of Fig.~\ref{peda-1-eta-pair}.  Regardless of the 
ATD and/or GCM scheme, we observe a reduction of the collective masses
for increasing $\eta$-values. This agrees well with previous results  
\cite{Rayner-UPRC-2014,Rayner-UEPJA-2014,Rayner-RaEPJA-2016,Robledo-Giulliani}
as well as with the inverse dependence of the collective masses 
on the square of the pairing gap \cite{proportional-1,proportional-2}. Such a 
reduction has a strong impact on the predicted t$_{SF}$ values.  For example, in 
the case of $^{243}$U and E$_{0}$ = 1.0 MeV, we have obtained
log$_{10}$ t$_{SF}$= 60.5530 s, log$_{10}$ t$_{SF}$= 54.9162 s
and log$_{10}$ t$_{SF}$= 49.7725 s within the ATD scheme 
for $\eta$ = 1.00, 1.05 and 1.10, respectively. The corresponding GCM values are 
log$_{10}$ t$_{SF}$= 39.2109 s, log$_{10}$ t$_{SF}$= 31.4145 s 
and log$_{10}$ t$_{SF}$= 24.9485 s. 

\begin{figure*}
\includegraphics[width=1.0\textwidth]{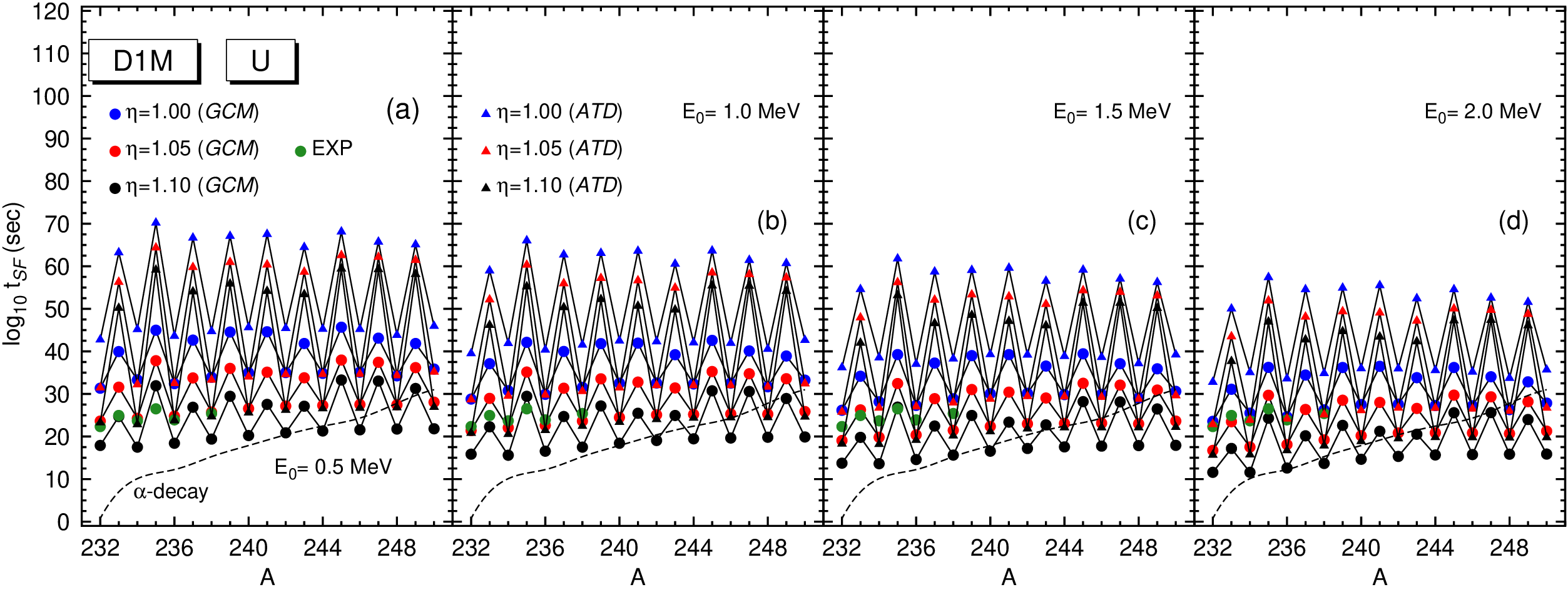}
\caption{(Color online) The spontaneous fission half-lives, predicted within the  
GCM  and ATD schemes, for the isotopes $^{232-250}$U are depicted 
as functions of the mass number A. Results have been obtained with the normal ($\eta$ =1.00) and modified 
($\eta$ =1.05 and 1.10) Gogny-D1M EDF. Calculations have been carried out with 
E$_{0}$ = 0.5 [panel (a)], 1.0 [panel (b)], 1.5 [panel (c)] and 
2.0 MeV [panel (d)], respectively. Results  for  even-even U isotopes  
are taken from Ref.~\cite{Rayner-UPRC-2014}. The experimental
t$_{SF}$ values are taken from Ref.~\cite{Holden-paper}. In addition, $\alpha$-decay half-lives 
are plotted with short dashed lines. For more details, see the main text.
}
\label{tsf-U-eta} 
\end{figure*}

\begin{figure*}
\includegraphics[width=1.0\textwidth]{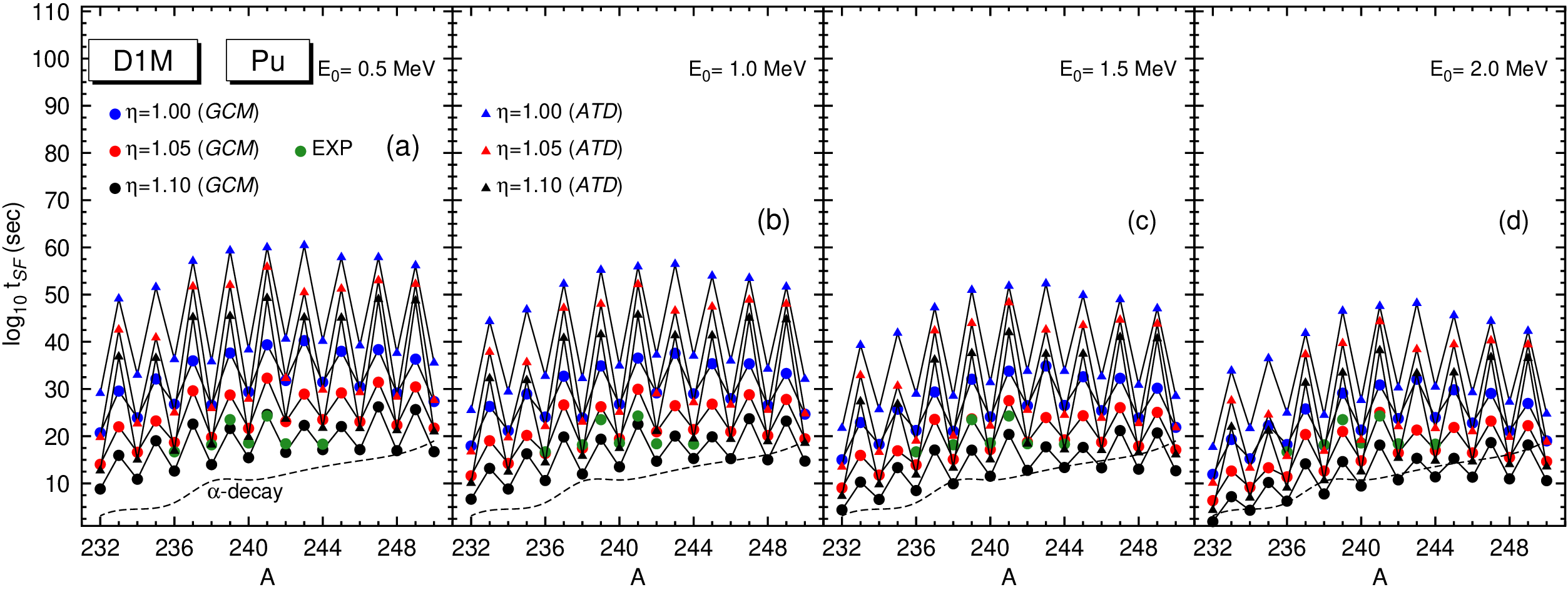}
\caption{(Color online) The spontaneous fission half-lives, predicted within the  
GCM  and ATD  schemes, for the isotopes $^{232-250}$Pu are depicted 
as functions of the mass number A. Results have been obtained with the normal ($\eta$ =1.00) and modified 
($\eta$ =1.05 and 1.10) Gogny-D1M EDF. Calculations have been carried out with 
E$_{0}$ = 0.5 [panel (a)], 1.0 [panel (b)], 1.5 [panel (c)] and 
2.0 MeV [panel (d)], respectively. Results  for  even-even Pu isotopes  
are taken from Ref.~\cite{Rayner-UEPJA-2014}. The experimental
t$_{SF}$ values are taken from Ref.~\cite{Holden-paper}. In addition, $\alpha$-decay half-lives 
are plotted with short dashed lines. For more details, see the main text. 
}
\label{tsf-Pu-eta} 
\end{figure*}

In Fig.~\ref{tsf-U-eta}  we have plotted,  the spontaneous fission half-lives, predicted within the  
GCM  and ATD  schemes, for U isotopes
 as functions of the mass number A.
The t$_{SF}$ values obtained for   Pu isotopes are shown in Fig.~\ref{tsf-Pu-eta} .
 Results have been obtained with
 the normal ($\eta$ =1.00) and modified 
($\eta$ =1.05 and 1.10) Gogny-D1M EDF. Calculations have been carried out with 
E$_{0}$ = 0.5 [panel (a))], 1.0 [panel (b)], 1.5 [panel (c)] and 
2.0 MeV [panel (d)], respectively. The experimental
t$_{SF}$ values are taken from Ref.~\cite{Holden-paper}. In addition, $\alpha$-decay half-lives 
are also included in the plots with short dashed lines.

As can be seen from Figs.~\ref{tsf-U-eta} and \ref{tsf-Pu-eta}, 
increasing the strengths of the pairing fields by 5 or 10 $\%$ leads to 
a pronounced reduction of several orders of magnitude in the predicted 
spontaneous fission half-lives. This is a consequence of the 
corresponding reduction in the ATD and/or GCM collective masses. Such a 
reduction in the predicted t$_{SF}$ values also tends to improve  the 
agreement with the available experimental data, especially within the 
GCM scheme. However, regardless of the $\eta$ value, the predicted  ATD 
spontaneous fission half-lives remain larger than the GCM ones. In the 
case of the odd-mass nuclei we have found, that the amplitude of the 
staggering does not exhibit a pronounced reduction as a function of 
$\eta$, with the ATD values being larger than the GCM ones. 
The amplitude of the staggering could be reduced in our calculation by
considering that dynamical pairing correlations are expected to be
larger for the odd isotopes than for the even ones. However, a qualitative
statement is difficult to asses until a detailed calculation including
particle number projection and fluctuations in the pairing gap are performed.
On the other hand, it is 
satisfying to see that, in spite of the large variability in the 
predicted t$_{SF}$ values with respect to the details involved in their 
computation, the main findings summarized in Fig.~\ref{tsf-U} still 
hold, i.e., regardless of the (ATD and/or GCM) scheme used as well as 
of  the considered E$_{0}$ and/or  $\eta$ values our Gogny-D1M HFB-EFA 
calculations predict larger t$_{SF}$ values for odd-mass U and Pu 
nuclei as compared with their even-even neighbors. On the other hand 
for both the U and Pu isotopic chains, we also observe a more 
pronounced competition between spontaneous fission and $\alpha$-decay 
with increasing mass  number A.

\section{Summary and conclusions}
\label{Coclusions}

In this paper, we have studied the fission properties of  odd-mass U 
and Pu nuclei. To this end, we have considered isotopes in the mass
range A=233,\ldots,249 as representative samples. 
We have resorted to the constrained  HFB-EFA in order to alleviate the 
already substantial computational effort required in the study of those 
odd-mass nuclear systems. Besides the usual constraints on both the 
proton $\hat{Z}$ and neutron $\hat{N}$ number operators, we have 
employed constraints on the axially symmetric quadrupole 
$\hat{Q}_{20}$, octupole $\hat{Q}_{30}$ and $\hat{Q}_{10}$ operators. 
On the other hand, HFB-EFA solutions corresponding to separated 
fragments have been reached with the help of constraints on the necking 
operator $\hat{Q}_{Neck}(z_{0},C_{0})$. We have presented a detailed 
account of the blocking methodology used to obtain 1F and 2F paths in 
the studied odd-mass nuclei. Zero-point quantum rotational and 
vibrational corrections have been added to the corresponding HFB-EFA 
energies {\it{a posteriori}}. The former has been computed in terms of 
the Yoccoz moment of inertia. On the other hand, both the GCM and ATD 
schemes have been used to obtain the collective masses and the 
vibrational corrections. 

The systematic of the fission paths shows a rich topology for odd-mass 
U and Pu nuclei similar to the one already found in previous studies 
for even-even systems in this region of the nuclear chart 
\cite{Rayner-UPRC-2014,Rayner-UEPJA-2014,Rayner-RaEPJA-2016}. Our 
Gogny-D1M calculations provide 1F paths with normal deformed, first and 
even second isomeric minima. In particular,  those second fission 
isomers show up around $^{239}$U and $^{243}$Pu. We conclude 
that the shell effects leading to such second fission isomers are 
systematically present in our calculations for even-even and odd-mass  
U and Pu nuclei. Furthermore, the 1F paths display first and  second 
barriers. Outer (third) barriers also emerge  along the 1F paths as we 
move towards the very neutron-rich sectors in both the U and Pu chains. 
All those quantities exhibit odd-even effects as functions of the mass 
number A. We have found that, for the considered odd-mass nuclei, the 
ground state (1F and 2F) fission paths are always higher than the AV 
ones. This is a manifestation of the specialization energy arising from 
the fact that we have followed configurations with a fixed K$_{0}$ 
value. Those specialization energy effects together with the quenching 
of pairing correlations, taken into account selfconsistently via the 
Ritz-variational solution of the HFB-EFA equations, lead to larger 
spontaneous fission half-lives, computed within the WKB approximation, 
in odd-mass U and Pu nuclei as compared with their even-even neighbors. 
We have found that $\alpha$-decay is the dominant decay channel for 
most of the studied U and Pu nuclei. However, the steady increase in 
the $\alpha$-decay lifetimes, as functions of the mass number A, leads 
to a pronounced competition with  spontaneous fission.

We have studied the masses, charges and shapes  of the fission fragments 
with the help of 2F HFB-EFA solutions at the largest quadrupole deformations. On the 
one hand, our results point to the key role played by the proton Z=50 and neutron
N=82 shell closures in determining the properties of the predicted fission products. On 
the other hand, for several of the studied odd-mass U and Pu nuclei, we have 
obtained oblate deformed fission fragments that deserve further attention as they
are usually assumed to be prolate. 

We have studied the impact of pairing correlations on the spontaneous 
fission half-lives obtained for the considered odd-mass U and Pu systems.
Our results, based on a modified Gogny-D1M EDF, reveal that increasing 
the strengths of the pairing fields by just 5 and 10 $\%$ lead to 
pronounced reductions of several orders of magnitude in the t$_{SF}$ values.
Those results and the fact that at least for some of the considered 
odd-mass nuclei we are dealing we a weak pairing regime, call for a more sophisticated 
treatment of the (spontaneously broken) U(1) particle number symmetry in which 
pairing fluctuations and their coupling to the relevant deformations are taken into 
account via the minimization of the action Eq. (\ref{action}) \cite{Action-Rayner}.
Nevertheless, in spite of the strong variance of the predicted 
fission rates with respect to the details involved in their 
computation, it is satisfying to observe  the robustness in the 
systematic of the predicted t$_{SF}$ values.

Finally, let us  mention that the results discussed in this paper 
represent a first step towards a description of the fission properties 
of even-even and odd-mass U and Pu nuclei on an equal footing.  A long 
list of tasks remains to be undertaken in the near future. For example, 
in addition to a more realistic treatment of pairing correlations, 
several aspects related with the computation of the collective masses 
as well as the impact of triaxiality in some sectors of the 1F paths 
obtained for odd-mass U and Pu nuclei remain to be clarified. Work along 
these lines is in progress and will be reported elsewhere.

\begin{acknowledgments}
The work of LMR was partly supported by  Spanish MINECO grant Nos
FPA2015-65929 and FIS2015-63770.

\end{acknowledgments}


\begin{thebibliography}{00}

\bibitem{Bjor} S. Bj\"ornholm and J.E. Lynn, Rev. Mod. Phys. {\bf{52}}, 725 (1980).

\bibitem{Specht} H.J. Specht, Rev. Mod. Phys. {\bf{46}}, 773 (1974).


\bibitem{Meitner} L. Meitner and O. R. Frish, Nature {\bf{143}}, 239 (1939).


\bibitem{Wagemans} C. Wagemans, {\it{ The Nuclear Fission Process}} (CRC Press, Boca 
Raton, 1991).



\bibitem{Baran-Kowal-others-review2015} A. Baran, M. Kowal, P. -G. Reinhard, 
L. M. Robledo, A. Staszczak and M. Warda, Nucl. Phys. A {\bf 994}, 442 (2015).


\bibitem{SR-Review-2016} N. Schunck and L. M. Robledo, Rep. Prog. Phys. {\bf{79}}, 116301 (2016).


\bibitem{Bender-1998} M. Bender, K. Rutz, P.-G. Reinhard, J.A. Maruhn 
and W. Greiner, Phys. Rev. C {\bf 58}, 2126 (1998). 


\bibitem{Warda-Egido-2012} 
M. Warda and J.L. Egido, 
Phys. Rev. C {\bf{86}}, 014322 (2012).


\bibitem{Warda-Egido-Robledo-Pomorski-2002} 
M. Warda, J. L. Egido, L.M. Robledo  and K. Pomorski, 
Phys. Rev. C {\bf 66}, 014310 (2002).


\bibitem{Nature-Naza} S. \'Cwiok, P. -H. Heenen and W. Nazarewicz, Nature {\bf{433}}, 705 (2005).


\bibitem{Viola-Seaborg} V.E. Viola Jr. and G.T. Seaborg, J. Inorg. Nucl. Chem. {\bf{28}}, 741 (1966).


\bibitem{TDong2005} T. Dong and Z. Ren, Eur. Phys. J. A {\bf{26}}, 69 (2005).
 
\bibitem{Rayner-UPRC-2014} R. Rodr\'iguez-Guzm\'an and 
L.M. Robledo, Phys. Rev. C {\bf{89}}, 054310 (2014). 

\bibitem{Rayner-UEPJA-2014} R. Rodr\'iguez-Guzm\'an and 
L.M. Robledo, Eur. Phys. J. A {\bf{50}}, 142 (2014). 

\bibitem{Rayner-RaEPJA-2016} R. Rodr\'iguez-Guzm\'an and 
L.M. Robledo, Eur. Phys. J. A {\bf{52}}, 12 (2016).


\bibitem{Rayner-No-odd} R. Rodr\'iguez-Guzm\'an and 
L.M. Robledo, Eur. Phys. J. A {\bf{52}}, 348 (2016).


\bibitem{Robledo-Giulliani} 
S. A. Giuliani and L.M Robledo, 
Phys. Rev. C \textbf{88}, 054325 (2013).

\bibitem{Cluster-Warda} M. Warda and L. M. Robledo, Phys. Rev. C {\bf{84}}, 044608 (2011).


\bibitem{Panov} I. V. Panov, I. Yu. Korneev and F. -K. Thielemann, Astron. Lett. {\bf{34}}, 189 (2008).

\bibitem{GMP_FR} G. Mart\'inez-Pinedo {\it{et al}}, Prog. in Part. and Nucl. Phys. {\bf{59}}, 199 (2007).


\bibitem{syst-SH-SGR} S. A. Giuliani, G. Mart\'inez-Pinedo and 
L. M. Robledo, arXiV/nucl-th/1704.00554 (2017).


\bibitem{Krappe} 
H.J. Krappe and K. Pomorski, 
{\it{Theory of Nuclear Fission}}, 
Lectures  Notes in Physics, {\bf{838}} (Springer, Berlin, 2012).


\bibitem{Sierk-PRC2015} P. M\"oller, A. J. Sierk, T. Ichikawa, A. Iwamoto
and M. Mumpower, Phys. Rev. C {\bf{91}}, 024310 (2015). 


\bibitem{JULIN-SHE} 
R. Julin, 
Nucl. Phys. A {\bf{834}}, 15c (2010).


\bibitem{rs} 
P. Ring and P. Schuck, 
{\em The Nuclear Many-Body Problem} 
(Springer, Berlin, 1980).

\bibitem{Action-Rayner} S. A. Giuliani, L.M Robledo and 
R. Rodr\'iguez-Guzm\'an, Phys. Rev. C {\bf{90}}, 054311 (2014). 


\bibitem{gogny-d1s} 
J. F. Berger, M. Girod, and D. Gogny, 
Nucl. Phys. A {\bf{428}}, 23c (1984).

\bibitem{Delaroche-2006} 
J.-P. Delaroche, M. Girod, H. Goutte and J. Libert, 
Nucl. Phys. A {\bf{771}}, 103 (2006).


\bibitem{Dubray} 
N. Dubray, H. Goutte and J.-P. Delaroche, 
Phys. Rev. C {\bf{77}}, 014310 (2008).

\bibitem{Younes-fission} W. Younes and D. Gogny, Phys. Rev. C {\bf{80}}, 054313 (2009).


\bibitem{UNEDF1} N. Nikolov, N. Schunck, W. Nazarewicz, M. Bender
 and J. Pei, Phys. Rev. C {\bf{83}}, 034305 (2011).
 
\bibitem{Mcdonell-2} 
J.D. McDonnell, W. Nazarewicz and J.A. Sheikh, 
Phys. Rev. C {\bf{87}}, 054327 (2013). 

\bibitem{Erler2012} 
J. Erler, K. Langanke, H.P. Loens, G. Mart\'inez-Pinedo and P.-G. Reinhard, 
Phys. Rev. C {\bf{85}}, 025802 (2012). 

\bibitem{Baran-SF-2012} 
A. Staszczak, A. Baran, W. Nazarewicz, 
Physical Review C \textbf{87}, 024320 (2013).

\bibitem{Baran-1981} 
A. Baran, K. Pomorski, A. Lukasiak and A. Sobiczewski, 
Nucl. Phys. A {\bf{361}}, 83 (1981).

\bibitem{BCPM}  
M. Baldo, L.M. Robledo, P. Schuck and X. Vi\~nas, 
Phys. Rev. C {\bf{87}}, 064305 (2013).

\bibitem{Abusara-2010} 
H. Abusara, A.V. Afanasjev and P. Ring, 
Phys. Rev. C {\bf{82}}, 044303 (2010).

\bibitem{Abu-2012-bheights} 
H. Abusara, A.V. Afanasjev and  P. Ring, 
Phys. Rev. C {\bf 85}, 024314 (2012).

\bibitem{RMF-LU-2012} 
B.-N. Lu, E.-G. Zhao and S.-G. Zhou, 
Phys. Rev. C {\bf{85}}, 011301 (2012).

\bibitem{Kara-RMF} 
S. Karatzikos, A. V. Afanasjev, G. A. Lalazissis and  P. Ring, 
Phys. Lett. B {\bf{689}}, 72 (2010).

\bibitem{Hamamoto-dd} I. Hamamoto, Phys. Rev. C {\bf{79}} 014307 (2009).

\bibitem{EFA-jsut} S. Perez-Martin and L.M. Robledo, Phys. Rev. C {\bf{78}}, 014304 (2008).

\bibitem{EFA-Bonneau} L. Bonneau, P. Quentin and P. M\"oller, Phys. Rev. C {\bf{76}}, 024320 (2007).

\bibitem{duguet-odd} T. Duguet, P. Bonche, P. -H. Heenen 
and J. Meyer, Phys. Rev. C {\bf{65}}, 014310 (2001).

\bibitem{EFA-Rayner-1} R. Rodr\'iguez-Guzm\'an, P. Sarriguren, L.M. Robledo
and S. Perez-Martin, Phys. Lett. B {\bf{691}}, 202 (2010).

\bibitem{EFA-Rayner-2} R. Rodr\'iguez-Guzm\'an, P. Sarriguren and L.M. Robledo, Phys. Rev. C {\bf{82}}, 044318 (2010).

\bibitem{EFA-Rayner-3} R. Rodr\'iguez-Guzm\'an, P. Sarriguren and L.M. Robledo, Phys. Rev. C {\bf{82}}, 061302(R) (2010).

\bibitem{EFA-Rayner-4} R. Rodr\'iguez-Guzm\'an, P. Sarriguren and L.M. Robledo, Phys. Rev. C {\bf{83}}, 044307 (2011).


\bibitem{reorient-1} N. Schunck et al., Phys. Rev. C {\bf{81}}, 024316 (2010).

\bibitem{reorient-2} P. Olbratowski, J. Dobaczewski, J. Dudek 
and W. Pl\'ociennik, Phys. Rev. Lett. {\bf{93}}, 052501 (2004).

\bibitem{EFA-Decharge} J. Decharg\'e and D. Gogny, Phys. Rev. C {\bf{21}}, 1568 (1980).


\bibitem{Robledo-Bertsch2OGM} 
L.M. Robledo and G. F. Berstch, 
Phys. Rev. C {\bf 84}, 014312 (2011).


\bibitem{Holden-paper} 
N. E. Holden and D. C. Hoffman, 
Pure Appl. Chem. {\bf{72}}, 1525 (2000).


\bibitem{special-Fong} P. Fong, Phys. Rev. C {\bf 122}, 1545 (1961).


\bibitem{proportional-1} 
M. Brack, J. Damgaard, A.S. Jensen, H.C. Pauli, V.M Strutinsky and C.Y. Wong, 
Rev. Mod. Phys. {\bf{44}}, 320 (1972).

\bibitem{proportional-2} 
J.F. Berstch and H. Flocard, 
Phys. Rev. C {\bf{43}}, 2200 (1991).






\bibitem{gogny-d1m} 
S. Goriely, S. Hilaire, M. Girod  and S. P\'eru, 
Phys. Rev. Lett. {\bf 102}, 242501 (2009).

\bibitem{gogny} 
J. Decharg\'e and D. Gogny, 
Phys. Rev. C {\bf 21}, 1568 (1980).

\bibitem{PRCQ2Q3-2012} 
R. Rodr\'iguez-Guzm\'an, L.M. Robledo and P. Sarriguren, 
Phys. Rev. C {\bf 86}, 034336 (2012).

\bibitem{Robledo-Rayner-JPG-2012}  
L.M. Robledo and  R. Rodr\'iguez-Guzm\'an, 
J. Phys. G: Nucl. Part. Phys. {\bf{39}}, 105103 (2012).

\bibitem{PTpaper-Rayner} 
R. Rodr\'iguez-Guzm\'an, L.M. Robledo, P. Sarriguren  and J. E. Garc\'ia-Ramos, 
Phys. Rev. C {\bf 81}, 024310 (2010).

\bibitem{Rayner-Robledo-JPG-2009} L.M. Robledo, R. Rodr\'iguez-Guzm\'an, and P. Sarriguren, J.
Phys. G: Nucl. Part. Phys. {\bf{36}}, 115104 (2009).

\bibitem{Gaudin-th} M. Gaudin, Nucl. Phys. {\bf{15}}, 89 (1960).

\bibitem{Gaudin-Sara} S. Perez-Martin and L.M. Robledo, Phys. Rev. C {\bf{76}}, 064314 (2007).

\bibitem{CoulombSlater} 
C. Titin-Schnaider and Ph. Quentin, Phys. Lett. B {\bf{49}}, 213 (1974).


\bibitem{RCORR-1} 
R. Rodr\'iguez-Guzm\'an, J.L. Egido and 
 L.M. Robledo, Phys. Lett. B {\bf{474}}, 15 (2000). 

\bibitem{RCORR-2} R. Rodr\'iguez-Guzm\'an, J.L. Egido 
and  L.M. Robledo, Phys. Rev. C {\bf 62}, 054308 (2000). 


\bibitem{RCORR-3} 
J.L. Egido and L.M.Robledo, 
Lectures Notes in Physics {\bf 641}, 269 (2004).


\bibitem{ATDHFB-T-1} M. Baranger and M. Veneroni, Ann. Phys. {\bf 114}, 123 (1978).

\bibitem{ATDHFB-T-2} H.M. Sommermann, Ann. Phys. {\bf 151}, 163 (1983).

\bibitem{ATDHFB-T-3} P. Ring, L.M. Robledo, J.L. Egido 
and M. Faber, Nucl. Phys. A {\bf 419}, 261 (1984).


\bibitem{crankingAPPROX} 
M. Girod and B. Grammaticos, 
Nucl. Phys. A {\bf{330}}, 40 (1979).

\bibitem{Giannoni-1} 
M.J. Giannoni  and P. Quentin, 
Phys. Rev. C {\bf{21}}, 2060 (1980). 

\bibitem{Giannoni-2} M.J. Giannoni  and P. Quentin,
Phys. Rev. C {\bf{21}}, 2076 (1980).

\bibitem{Libert-1999} 
J. Libert, M. Girod and  J.P. Delaroche,  
Phys. Rev. C {\bf{60}}, 054301 (1999).


\bibitem{Baran-TSF-1} 
A. Baran, 
Phys. Lett. B {\bf{76}}, 8 (1978).

\bibitem{Baran-TSF-2} 
A. Baran, J. A. Sheikh, J. Dobaczewski, W. Nazarewicz  and A. Staszczak, 
Phys. Rev. C {\bf{84}}, 054321 (2011).

\bibitem{Kowal-3fi-1} M. Kowal and J. Skalski, Phys. Rev. C {\bf{85}}, 061302(R)
(2012).

\bibitem{Kowal-3fi-2} P. Jachimowicz, M. Kowal and J. Skalski, Phys. Rev. C {\bf{87}}, 044308 
(2013).

\bibitem{Pas-3fi-1} V. V. Pashkevich, Nucl. Phys. A {\bf{169}}, 275 (1971).

\bibitem{Moeller-3fi-1} P. M\"oller, Nucl. Phys. A {\bf{192}}, 529 (1972).

\bibitem{Mahrun-3fi-1} K. Rutz, J. Mahrun, P. -G. Reinhard 
and W. Greiner, Nucl. Phys. A {\bf{590}}, 680 (1995).

\bibitem{Berger-3fi-1} J. F. Berger, M. Girod and D. Gogny, Nucl. Phys. A {\bf{502}}, 85 (1989).

\bibitem{Zhao-3fi-1} J. Zhao, B. -N. Lu, D. Vretenar, E. -G. Zhao and  S. -G. Zhou, Phys. Rev. C {\bf{91}}, 014321 (2015).

\bibitem{Nenoff-2007} N. Nenoff, P. Bringel, A. B\"urger, S Chmel, S. Dababneh, M. Heil, H. H\"ubel, F. K\"appeler, A. Neusser-Neffgen
and R. Plag, Eur. Phys. J. A {\bf{32}}, 165 (2007).
  


\bibitem{Ter-1996} G. M. Ter-Akopian, J. H. Hamilton, Yu. Ts. Oganessian, A. V. Daniel, J. Kormicki, A. V. Ramayya, 
G. S. Popeko, B. R. S. Babu, Q.-H. Lu, K. Butler-Moore, W. -C. Ma, S. \'Cwiok, W. Nazarewicz, J. K. Deng, D. Shi,
J. Kliman, M. Morhac, J. D. Cole, R. Aryaeinejad, N. R. Johnson, I. Y. Lee, F. K. McGowan and J. X. Saladin, Phys. 
Rev. Lett. {\bf{77}}, 32 (1996).

\bibitem{Piessens-1993} M. Piessens, E. Jacobs, S. Pomm\'e and D. D. Frenne, Nucl. Phys. A {\bf{556}}, 88 (1993).


\bibitem{Pu-mass-fragments-exp-1} L. Dematt\'e, C. Wagemans, R. Barth\'el\'emy, R. D\'hont
and A. Deruytter, Nucl. Phys. A {\bf{617}}, 331 (1997).

\bibitem{Pu-mass-fragments-exp-2} D.C. Hoffman and 
M.M. Hoffman, Ann. Rev. Nucl. Sci. {\bf{24}}, 151 (1974).

\bibitem{Goutte-dynamical-distribution} H. Goutte, J. F. Berger, P. Casoli and D. Gogny,  Phys. Rev. C {\bf{71}}, 024316 (2005).

\bibitem{Chasman-breaking} B. D. Wilkins, E. P. Steinberg and R. R. Chasman, Phys. Rev. C {\bf{14}}, 1832 (1976).

\bibitem{EXP-Spins-Gs-U-Pu} Brookhaven National Nuclear Data Center, http://www.nndc.bnl.gov/nudat2

\bibitem{Schmidt-fragments} K.-H. Schmidt et al., Nucl. Phys. A {\bf{665}}, 221 (2000). 

\bibitem{Moller-1} 
P. M\"oller and A. Iwamoto, 
Phys. Rev. C {\bf{61}}, 047602 (2000).

\bibitem{Moller-2} 
P. M\"oller, D.G. Madlan, A.J. Sierk and A. Iwamoto, 
Nature {\bf{409}}, 785 (2001).

\end{thebibliography}
\end{document}